\def\gax{\mathrel{\raise.3ex\hbox{$>$}\mkern-14mu\lower0.6ex\hbox{$\sim$}}}
\def\lax{\mathrel{\raise.3ex\hbox{$<$}\mkern-14mu\lower0.6ex\hbox{$\sim$}}}
\def\etal{{~et~al.}}

\def\VEV#1{\left\langle #1\right\rangle}

\def\kms{{\rm\,km\,s^{-1}}}

\documentstyle[11pt,aaspp4,flushrt]{article}
\def\PsfigVersion{1.10}
\def\setDriver{\DvipsDriver} 
\ifx\undefined\psfig\else \fi
\let\LaTeXAtSign=\@
\let\@=\relax
\edef\psfigRestoreAt{\catcode`\@=\number\catcode`@\relax}
\catcode`\@=11\relax
\newwrite\@unused
\def\ps@typeout#1{{\let\protect\string\immediate\write\@unused{#1}}}

\def\DvipsDriver{
	\ps@typeout{psfig/tex \PsfigVersion -dvips}
\def\PsfigSpecials{\DvipsSpecials} 	\def\ps@dir{/}
\def\ps@predir{} }
\def\OzTeXDriver{
	\ps@typeout{psfig/tex \PsfigVersion -oztex}
	\def\PsfigSpecials{\OzTeXSpecials}
	\def\ps@dir{:}
	\def\ps@predir{:}
	\catcode`\^^J=5
}

\def\figurepath{./:}

\def\DoPaths#1{\expandafter\EachPath#1\stoplist}
\def\leer{}
\def\EachPath#1:#2\stoplist{
  \ExistsFile{#1}{\SearchedFile}
  \ifx#2\leer
  \else
    \expandafter\EachPath#2\stoplist
  \fi}
\def\ps@dir{/}
\def\ExistsFile#1#2{%
   \openin1=\ps@predir#1\ps@dir#2
   \ifeof1
       \closein1
   \else
       \closein1
        \ifx\ps@founddir\leer
           \edef\ps@founddir{#1}
        \fi
   \fi}
\def\get@dir#1{%
  \def\ps@founddir{}
  \def\SearchedFile{#1}
  \DoPaths\figurepath
}

\def\@nnil{\@nil}
\def\@empty{}
\def\@psdonoop#1\@@#2#3{}
\def\@psdo#1:=#2\do#3{\edef\@psdotmp{#2}\ifx\@psdotmp\@empty \else
    \expandafter\@psdoloop#2,\@nil,\@nil\@@#1{#3}\fi}
\def\@psdoloop#1,#2,#3\@@#4#5{\def#4{#1}\ifx #4\@nnil \else
       #5\def#4{#2}\ifx #4\@nnil \else#5\@ipsdoloop #3\@@#4{#5}\fi\fi}
\def\@ipsdoloop#1,#2\@@#3#4{\def#3{#1}\ifx #3\@nnil 
       \let\@nextwhile=\@psdonoop \else
      #4\relax\let\@nextwhile=\@ipsdoloop\fi\@nextwhile#2\@@#3{#4}}
\def\@tpsdo#1:=#2\do#3{\xdef\@psdotmp{#2}\ifx\@psdotmp\@empty \else
    \@tpsdoloop#2\@nil\@nil\@@#1{#3}\fi}
\def\@tpsdoloop#1#2\@@#3#4{\def#3{#1}\ifx #3\@nnil 
       \let\@nextwhile=\@psdonoop \else
      #4\relax\let\@nextwhile=\@tpsdoloop\fi\@nextwhile#2\@@#3{#4}}
\ifx\undefined\fbox
\newdimen\fboxrule
\newdimen\fboxsep
\newdimen\ps@tempdima
\newbox\ps@tempboxa
\long\def\fbox#1{\leavevmode\setbox\ps@tempboxa\hbox{#1}\ps@tempdima\fboxrule
    \advance\ps@tempdima \fboxsep \advance\ps@tempdima \dp\ps@tempboxa
   \hbox{\lower \ps@tempdima\hbox
  {\vbox{\hrule height \fboxrule
          \hbox{\vrule width \fboxrule \hskip\fboxsep
          \vbox{\vskip\fboxsep \box\ps@tempboxa\vskip\fboxsep}\hskip 
                 \fboxsep\vrule width \fboxrule}
                 \hrule height \fboxrule}}}}
\fi
\newread\ps@stream
\newif\ifnot@eof       
\newif\if@noisy        
\newif\if@atend        
\newif\if@psfile       
{\catcode`\%=12\global\gdef\epsf@start{
\def\epsf@PS{PS}
\def\epsf@getbb#1{%
\openin\ps@stream=\ps@predir#1
\ifeof\ps@stream\ps@typeout{Error, File #1 not found}\else
   {\not@eoftrue \chardef\other=12
    \def\do##1{\catcode`##1=\other}\dospecials \catcode`\ =10
    \loop
       \if@psfile
	  \read\ps@stream to \epsf@fileline
       \else{
	  \obeyspaces
          \read\ps@stream to \epsf@tmp\global\let\epsf@fileline\epsf@tmp}
       \fi
       \ifeof\ps@stream\not@eoffalse\else
       \if@psfile\else
       \expandafter\epsf@test\epsf@fileline:. \\%
       \fi
          \expandafter\epsf@aux\epsf@fileline:. \\%
       \fi
   \ifnot@eof\repeat
   }\closein\ps@stream\fi}%
\long\def\epsf@test#1#2#3:#4\\{\def\epsf@testit{#1#2}
			\ifx\epsf@testit\epsf@start\else
\ps@typeout{Warning! File does not start with `\epsf@start'.  It may not be a PostScript file.}
			\fi
			\@psfiletrue} 
{\catcode`\%=12\global\let\epsf@percent=
\long\def\epsf@aux#1#2:#3\\{\ifx#1\epsf@percent
   \def\epsf@testit{#2}\ifx\epsf@testit\epsf@bblit
	\@atendfalse
        \epsf@atend #3 . \\%
	\if@atend	
	   \if@verbose{
		\ps@typeout{psfig: found `(atend)'; continuing search}
	   }\fi
        \else
        \epsf@grab #3 . . . \\%
        \not@eoffalse
        \global\no@bbfalse
        \fi
   \fi\fi}%
\def\epsf@grab #1 #2 #3 #4 #5\\{%
   \global\def\epsf@llx{#1}\ifx\epsf@llx\empty
      \epsf@grab #2 #3 #4 #5 .\\\else
   \global\def\epsf@lly{#2}%
   \global\def\epsf@urx{#3}\global\def\epsf@ury{#4}\fi}%
\def\epsf@atendlit{(atend)} 
\def\epsf@atend #1 #2 #3\\{%
   \def\epsf@tmp{#1}\ifx\epsf@tmp\empty
      \epsf@atend #2 #3 .\\\else
   \ifx\epsf@tmp\epsf@atendlit\@atendtrue\fi\fi}

\chardef\psletter = 11 
\chardef\other = 12

\newif \ifdebug 
\newif\ifc@mpute 
\c@mputetrue 

\let\then = \relax
\def\r@dian{pt }
\let\r@dians = \r@dian
\let\dimensionless@nit = \r@dian
\let\dimensionless@nits = \dimensionless@nit
\def\internal@nit{sp }
\let\internal@nits = \internal@nit
\newif\ifstillc@nverging
\def \Mess@ge #1{\ifdebug \then \message {#1} \fi}

{ 
	\catcode `\@ = \psletter
	\gdef \nodimen {\expandafter \n@dimen \the \dimen}
	\gdef \term #1 #2 #3%
	       {\edef \t@ {\the #1}
		\edef \t@@ {\expandafter \n@dimen \the #2\r@dian}%
		\t@rm {\t@} {\t@@} {#3}%
	       }
	\gdef \t@rm #1 #2 #3%
	       {{%
		\count 0 = 0
		\dimen 0 = 1 \dimensionless@nit
		\dimen 2 = #2\relax
		\Mess@ge {Calculating term #1 of \nodimen 2}%
		\loop
		\ifnum	\count 0 < #1
		\then	\advance \count 0 by 1
			\Mess@ge {Iteration \the \count 0 \space}%
			\Multiply \dimen 0 by {\dimen 2}%
			\Mess@ge {After multiplication, term = \nodimen 0}%
			\Divide \dimen 0 by {\count 0}%
			\Mess@ge {After division, term = \nodimen 0}%
		\repeat
		\Mess@ge {Final value for term #1 of 
				\nodimen 2 \space is \nodimen 0}%
		\xdef \Term {#3 = \nodimen 0 \r@dians}%
		\aftergroup \Term
	       }}
	\catcode `\p = \other
	\catcode `\t = \other
	\gdef \n@dimen #1pt{#1} 
}

\def \Divide #1by #2{\divide #1 by #2} 

\def \Multiply #1by #2
       {{
	\count 0 = #1\relax
	\count 2 = #2\relax
	\count 4 = 65536
	\Mess@ge {Before scaling, count 0 = \the \count 0 \space and
			count 2 = \the \count 2}%
	\ifnum	\count 0 > 32767 
	\then	\divide \count 0 by 4
		\divide \count 4 by 4
	\else	\ifnum	\count 0 < -32767
		\then	\divide \count 0 by 4
			\divide \count 4 by 4
		\else
		\fi
	\fi
	\ifnum	\count 2 > 32767 
	\then	\divide \count 2 by 4
		\divide \count 4 by 4
	\else	\ifnum	\count 2 < -32767
		\then	\divide \count 2 by 4
			\divide \count 4 by 4
		\else
		\fi
	\fi
	\multiply \count 0 by \count 2
	\divide \count 0 by \count 4
	\xdef \product {#1 = \the \count 0 \internal@nits}%
	\aftergroup \product
       }}

\def\r@duce{\ifdim\dimen0 > 90\r@dian \then   
		\multiply\dimen0 by -1
		\advance\dimen0 by 180\r@dian
		\r@duce
	    \else \ifdim\dimen0 < -90\r@dian \then  
		\advance\dimen0 by 360\r@dian
		\r@duce
		\fi
	    \fi}

\def\Sine#1%
       {{%
	\dimen 0 = #1 \r@dian
	\r@duce
	\ifdim\dimen0 = -90\r@dian \then
	   \dimen4 = -1\r@dian
	   \c@mputefalse
	\fi
	\ifdim\dimen0 = 90\r@dian \then
	   \dimen4 = 1\r@dian
	   \c@mputefalse
	\fi
	\ifdim\dimen0 = 0\r@dian \then
	   \dimen4 = 0\r@dian
	   \c@mputefalse
	\fi
	\ifc@mpute \then
		\divide\dimen0 by 180
		\dimen0=3.141592654\dimen0
		\dimen 2 = 3.1415926535897963\r@dian 
		\divide\dimen 2 by 2 
		\Mess@ge {Sin: calculating Sin of \nodimen 0}%
		\count 0 = 1 
		\dimen 2 = 1 \r@dian 
		\dimen 4 = 0 \r@dian 
		\loop
			\ifnum	\dimen 2 = 0 
			\then	\stillc@nvergingfalse 
			\else	\stillc@nvergingtrue
			\fi
			\ifstillc@nverging 
			\then	\term {\count 0} {\dimen 0} {\dimen 2}%
				\advance \count 0 by 2
				\count 2 = \count 0
				\divide \count 2 by 2
				\ifodd	\count 2 
				\then	\advance \dimen 4 by \dimen 2
				\else	\advance \dimen 4 by -\dimen 2
				\fi
		\repeat
	\fi		
			\xdef \sine {\nodimen 4}%
       }}

\def\Cosine#1{\ifx\sine\UnDefined\edef\Savesine{\relax}\else
		             \edef\Savesine{\sine}\fi
	{\dimen0=#1\r@dian\advance\dimen0 by 90\r@dian
	 \Sine{\nodimen 0}
	 \xdef\cosine{\sine}
	 \xdef\sine{\Savesine}}}	      

\def\psdraft{
	\def\@psdraft{0}
}
\def\psfull{
	\def\@psdraft{100}
}

\psfull

\newif\if@scalefirst
\def\psscalefirst{\@scalefirsttrue}
\def\psrotatefirst{\@scalefirstfalse}
\psrotatefirst

\newif\if@draftbox
\def\psnodraftbox{
	\@draftboxfalse
}
\def\psdraftbox{
	\@draftboxtrue
}
\@draftboxtrue

\newif\if@prologfile
\newif\if@postlogfile
\def\pssilent{
	\@noisyfalse
}
\def\psnoisy{
	\@noisytrue
}
\psnoisy
\newif\if@bbllx
\newif\if@bblly
\newif\if@bburx
\newif\if@bbury
\newif\if@height
\newif\if@width
\newif\if@rheight
\newif\if@rwidth
\newif\if@angle
\newif\if@clip
\newif\if@verbose
\def\@p@@sclip#1{\@cliptrue}
\newif\if@decmpr
\def\@p@@sfigure#1{\def\@p@sfile{null}\def\@p@sbbfile{null}\@decmprfalse
   \openin1=\ps@predir#1
   \ifeof1
	\closein1
	\get@dir{#1}
	\ifx\ps@founddir\leer
		\openin1=\ps@predir#1.bb
		\ifeof1
			\closein1
			\get@dir{#1.bb}
			\ifx\ps@founddir\leer
				\ps@typeout{Can't find #1 in \figurepath}
			\else
				\@decmprtrue
				\def\@p@sfile{\ps@founddir\ps@dir#1}
				\def\@p@sbbfile{\ps@founddir\ps@dir#1.bb}
			\fi
		\else
			\closein1
			\@decmprtrue
			\def\@p@sfile{#1}
			\def\@p@sbbfile{#1.bb}
		\fi
	\else
		\def\@p@sfile{\ps@founddir\ps@dir#1}
		\def\@p@sbbfile{\ps@founddir\ps@dir#1}
	\fi
   \else
	\closein1
	\def\@p@sfile{#1}
	\def\@p@sbbfile{#1}
   \fi
}
\def\@p@@sfile#1{\@p@@sfigure{#1}}
\def\@p@@sbbllx#1{
		\@bbllxtrue
		\dimen100=#1
		\edef\@p@sbbllx{\number\dimen100}
}
\def\@p@@sbblly#1{
		\@bbllytrue
		\dimen100=#1
		\edef\@p@sbblly{\number\dimen100}
}
\def\@p@@sbburx#1{
		\@bburxtrue
		\dimen100=#1
		\edef\@p@sbburx{\number\dimen100}
}
\def\@p@@sbbury#1{
		\@bburytrue
		\dimen100=#1
		\edef\@p@sbbury{\number\dimen100}
}
\def\@p@@sheight#1{
		\@heighttrue
		\dimen100=#1
   		\edef\@p@sheight{\number\dimen100}
}
\def\@p@@swidth#1{
		\@widthtrue
		\dimen100=#1
		\edef\@p@swidth{\number\dimen100}
}
\def\@p@@srheight#1{
		\@rheighttrue
		\dimen100=#1
		\edef\@p@srheight{\number\dimen100}
}
\def\@p@@srwidth#1{
		\@rwidthtrue
		\dimen100=#1
		\edef\@p@srwidth{\number\dimen100}
}
\def\@p@@sangle#1{
		\@angletrue
		\edef\@p@sangle{#1} 
}
\def\@p@@ssilent#1{ 
		\@verbosefalse
}
\def\@p@@sprolog#1{\@prologfiletrue\def\@prologfileval{#1}}
\def\@p@@spostlog#1{\@postlogfiletrue\def\@postlogfileval{#1}}
\def\@cs@name#1{\csname #1\endcsname}
\def\@setparms#1=#2,{\@cs@name{@p@@s#1}{#2}}
\def\ps@init@parms{
		\@bbllxfalse \@bbllyfalse
		\@bburxfalse \@bburyfalse
		\@heightfalse \@widthfalse
		\@rheightfalse \@rwidthfalse
		\def\@p@sbbllx{}\def\@p@sbblly{}
		\def\@p@sbburx{}\def\@p@sbbury{}
		\def\@p@sheight{}\def\@p@swidth{}
		\def\@p@srheight{}\def\@p@srwidth{}
		\def\@p@sangle{0}
		\def\@p@sfile{} \def\@p@sbbfile{}
		\def\@p@scost{10}
		\def\@sc{}
		\@prologfilefalse
		\@postlogfilefalse
		\@clipfalse
		\if@noisy
			\@verbosetrue
		\else
			\@verbosefalse
		\fi
}
\def\parse@ps@parms#1{
	 	\@psdo\@psfiga:=#1\do
		   {\expandafter\@setparms\@psfiga,}}
\newif\ifno@bb
\def\bb@missing{
	\if@verbose{
		\ps@typeout{psfig: searching \@p@sbbfile \space  for bounding box}
	}\fi
	\no@bbtrue
	\epsf@getbb{\@p@sbbfile}
        \ifno@bb \else \bb@cull\epsf@llx\epsf@lly\epsf@urx\epsf@ury\fi
}	
\def\bb@cull#1#2#3#4{
	\dimen100=#1 bp\edef\@p@sbbllx{\number\dimen100}
	\dimen100=#2 bp\edef\@p@sbblly{\number\dimen100}
	\dimen100=#3 bp\edef\@p@sbburx{\number\dimen100}
	\dimen100=#4 bp\edef\@p@sbbury{\number\dimen100}
	\no@bbfalse
}
\newdimen\p@intvaluex
\newdimen\p@intvaluey
\def\rotate@#1#2{{\dimen0=#1 sp\dimen1=#2 sp
		  \global\p@intvaluex=\cosine\dimen0
		  \dimen3=\sine\dimen1
		  \global\advance\p@intvaluex by -\dimen3
		  \global\p@intvaluey=\sine\dimen0
		  \dimen3=\cosine\dimen1
		  \global\advance\p@intvaluey by \dimen3
		  }}
\def\compute@bb{
		\no@bbfalse
		\if@bbllx \else \no@bbtrue \fi
		\if@bblly \else \no@bbtrue \fi
		\if@bburx \else \no@bbtrue \fi
		\if@bbury \else \no@bbtrue \fi
		\ifno@bb \bb@missing \fi
		\ifno@bb \ps@typeout{FATAL ERROR: no bb supplied or found}
			\no-bb-error
		\fi
		\count203=\@p@sbburx
		\count204=\@p@sbbury
		\advance\count203 by -\@p@sbbllx
		\advance\count204 by -\@p@sbblly
		\edef\ps@bbw{\number\count203}
		\edef\ps@bbh{\number\count204}
		\if@angle 
			\Sine{\@p@sangle}\Cosine{\@p@sangle}
	        	{\dimen100=\maxdimen\xdef\r@p@sbbllx{\number\dimen100}
					    \xdef\r@p@sbblly{\number\dimen100}
			                    \xdef\r@p@sbburx{-\number\dimen100}
					    \xdef\r@p@sbbury{-\number\dimen100}}
                        \def\minmaxtest{
			   \ifnum\number\p@intvaluex<\r@p@sbbllx
			      \xdef\r@p@sbbllx{\number\p@intvaluex}\fi
			   \ifnum\number\p@intvaluex>\r@p@sbburx
			      \xdef\r@p@sbburx{\number\p@intvaluex}\fi
			   \ifnum\number\p@intvaluey<\r@p@sbblly
			      \xdef\r@p@sbblly{\number\p@intvaluey}\fi
			   \ifnum\number\p@intvaluey>\r@p@sbbury
			      \xdef\r@p@sbbury{\number\p@intvaluey}\fi
			   }
			\rotate@{\@p@sbbllx}{\@p@sbblly}
			\minmaxtest
			\rotate@{\@p@sbbllx}{\@p@sbbury}
			\minmaxtest
			\rotate@{\@p@sbburx}{\@p@sbblly}
			\minmaxtest
			\rotate@{\@p@sbburx}{\@p@sbbury}
			\minmaxtest
			\edef\@p@sbbllx{\r@p@sbbllx}\edef\@p@sbblly{\r@p@sbblly}
			\edef\@p@sbburx{\r@p@sbburx}\edef\@p@sbbury{\r@p@sbbury}
		\fi
		\count203=\@p@sbburx
		\count204=\@p@sbbury
		\advance\count203 by -\@p@sbbllx
		\advance\count204 by -\@p@sbblly
		\edef\@bbw{\number\count203}
		\edef\@bbh{\number\count204}
}
\def\in@hundreds#1#2#3{\count240=#2 \count241=#3
		     \count100=\count240	
		     \divide\count100 by \count241
		     \count101=\count100
		     \multiply\count101 by \count241
		     \advance\count240 by -\count101
		     \multiply\count240 by 10
		     \count101=\count240	
		     \divide\count101 by \count241
		     \count102=\count101
		     \multiply\count102 by \count241
		     \advance\count240 by -\count102
		     \multiply\count240 by 10
		     \count102=\count240	
		     \divide\count102 by \count241
		     \count200=#1\count205=0
		     \count201=\count200
			\multiply\count201 by \count100
		 	\advance\count205 by \count201
		     \count201=\count200
			\divide\count201 by 10
			\multiply\count201 by \count101
			\advance\count205 by \count201
		     \count201=\count200
			\divide\count201 by 100
			\multiply\count201 by \count102
			\advance\count205 by \count201
		     \edef\@result{\number\count205}
}
\def\compute@wfromh{
		\in@hundreds{\@p@sheight}{\@bbw}{\@bbh}
		\edef\@p@swidth{\@result}
}
\def\compute@hfromw{
	        \in@hundreds{\@p@swidth}{\@bbh}{\@bbw}
		\edef\@p@sheight{\@result}
}
\def\compute@handw{
		\if@height 
			\if@width
			\else
				\compute@wfromh
			\fi
		\else 
			\if@width
				\compute@hfromw
			\else
				\edef\@p@sheight{\@bbh}
				\edef\@p@swidth{\@bbw}
			\fi
		\fi
}
\def\compute@resv{
		\if@rheight \else \edef\@p@srheight{\@p@sheight} \fi
		\if@rwidth \else \edef\@p@srwidth{\@p@swidth} \fi
}
\def\compute@sizes{
	\compute@bb
	\if@scalefirst\if@angle
	\if@width
	   \in@hundreds{\@p@swidth}{\@bbw}{\ps@bbw}
	   \edef\@p@swidth{\@result}
	\fi
	\if@height
	   \in@hundreds{\@p@sheight}{\@bbh}{\ps@bbh}
	   \edef\@p@sheight{\@result}
	\fi
	\fi\fi
	\compute@handw
	\compute@resv}
\def\OzTeXSpecials{
	\special{empty.ps /@isp {true} def}
	\special{empty.ps \@p@swidth \space \@p@sheight \space
			\@p@sbbllx \space \@p@sbblly \space
			\@p@sbburx \space \@p@sbbury \space
			startTexFig \space }
	\if@clip{
		\if@verbose{
			\ps@typeout{(clip)}
		}\fi
		\special{empty.ps doclip \space }
	}\fi
	\if@angle{
		\if@verbose{
			\ps@typeout{(rotate)}
		}\fi
		\special {empty.ps \@p@sangle \space rotate \space} 
	}\fi
	\if@prologfile
	    \special{\@prologfileval \space } \fi
	\if@decmpr{
		\if@verbose{
			\ps@typeout{psfig: Compression not available
			in OzTeX version \space }
		}\fi
	}\else{
		\if@verbose{
			\ps@typeout{psfig: including \@p@sfile \space }
		}\fi
		\special{epsf=\ps@predir\@p@sfile \space }
	}\fi
	\if@postlogfile
	    \special{\@postlogfileval \space } \fi
	\special{empty.ps /@isp {false} def}
}
\def\DvipsSpecials{
	\special{ps::[begin] 	\@p@swidth \space \@p@sheight \space
			\@p@sbbllx \space \@p@sbblly \space
			\@p@sbburx \space \@p@sbbury \space
			startTexFig \space }
	\if@clip{
		\if@verbose{
			\ps@typeout{(clip)}
		}\fi
		\special{ps:: doclip \space }
	}\fi
	\if@angle
		\if@verbose{
			\ps@typeout{(clip)}
		}\fi
		\special {ps:: \@p@sangle \space rotate \space} 
	\fi
	\if@prologfile
	    \special{ps: plotfile \@prologfileval \space } \fi
	\if@decmpr{
		\if@verbose{
			\ps@typeout{psfig: including \@p@sfile.Z \space }
		}\fi
		\special{ps: plotfile "`zcat \@p@sfile.Z" \space }
	}\else{
		\if@verbose{
			\ps@typeout{psfig: including \@p@sfile \space }
		}\fi
		\special{ps: plotfile \@p@sfile \space }
	}\fi
	\if@postlogfile
	    \special{ps: plotfile \@postlogfileval \space } \fi
	\special{ps::[end] endTexFig \space }
}
\def\psfig#1{\vbox {
	\ps@init@parms
	\parse@ps@parms{#1}
	\compute@sizes
	\ifnum\@p@scost<\@psdraft{
		\PsfigSpecials 
		\vbox to \@p@srheight sp{
			\hbox to \@p@srwidth sp{
				\hss
			}
		\vss
		}
	}\else{
		\if@draftbox{		
			\hbox{\fbox{\vbox to \@p@srheight sp{
			\vss
			\hbox to \@p@srwidth sp{ \hss 
			 \hss }
			\vss
			}}}
		}\else{
			\vbox to \@p@srheight sp{
			\vss
			\hbox to \@p@srwidth sp{\hss}
			\vss
			}
		}\fi

	}\fi
}}
\psfigRestoreAt
\setDriver
\let\@=\LaTeXAtSign

\begin{document}
\title{An Efficient Search for Gravitationally-Lensed Radio Lobes}

\centerline{\it Submitted to ApJ: 2000.07.28, revised 2000.09.12.}

\author{   J.~Leh\'ar\altaffilmark{1},
           A.~Buchalter\altaffilmark{2},
           R.G.~McMahon\altaffilmark{3},
           C.S.~Kochanek\altaffilmark{1},
           T.W.B.~Muxlow\altaffilmark{4}
       }
\altaffiltext{1}{ Center for Astrophysics,
                  60 Garden St, Cambridge, MA 02138, USA,
                  jlehar@cfa.harvard.edu. }
\altaffiltext{2}{ Theoretical Astrophysics Group, Caltech, 130-33, Pasadena, CA 91125,
                  ari@tapir.caltech.edu }
\altaffiltext{3}{ Institute of Astronomy,
                  Madingley Rd, Cambridge, CB3 0HA, UK }
\altaffiltext{4}{ Nuffield Radio Astronomy Laboratories,
                  Jodrell Bank, Cheshire SK11 9DL, UK }

\begin{abstract}
We performed an automated comparison
of the FIRST radio survey with the APM optical catalog 
to find radio lobes with optical counterparts. 
Based on an initial survey covering $\sim$3000 square degrees,
we selected a sample of 33 lens candidates for VLA confirmation.
VLA and optical observations of these candidates
yielded two lens systems, one a new discovery (J0816+5003),
and one of which was previously known (J1549+3047).
Two other candidates have radio lobes with galaxies superposed,
but lack evidence of multiple imaging. 
One of our targets (J0958+2947) is a projected close pair
of quasars (8$''$ separation at redshifts 2.064 and 2.744). 
Our search method is highly efficient, with $>$5\% 
of our observing targets being lensed, compared to the 
usual success rate of $<$1\%. 
Using the whole FIRST survey, we expect to find 5--10 lenses in short 
order using this approach, and the sample could increase to hundreds
of lensed lobes in the Northern sky, using deeper optical surveys and planned
upgrades to the VLA.
Such a sample would be a powerful probe of galaxy structure and evolution.
\end{abstract}


\section{Introduction}

The 60 known gravitational lenses are powerful astrophysical probes.
They are the only galaxies selected on mass
rather than luminosity, making them a unique sample with very
different properties and biases from traditional luminosity-selected
galaxies.  All lenses have extraordinarily precise
mass measurements (better than 5\% including systematics), allowing
detailed comparisons of galaxies of constant mass.
Lensing  has been used to constrain 
cosmological parameters (see reviews by Schneider\etal\ 1992;
Blandford \&~Narayan 1992), the structure and evolution of
galaxies (e.g., Keeton\etal\ 1997, Keeton\etal\ 1998, Kochanek\etal\ 2000),
quasar and AGN host galaxies (e.g., Keeton\etal\ 2000), and even the
interstellar medium and extinction in intermediate-redshift galaxies
(e.g., Falco\etal\ 1999).

Such astrophysical studies are limited by the sample size.
The properties of each individual lens depend on many
variables (the source redshift, the lens redshift, and the galaxy mass).
With only 60 lenses,
it is difficult to subdivide the sample over more than one variable.
For example, one cannot
separately explore the evolution of high and low mass galaxies, or the
average mass or number density of the lenses in each bin.
Only 10--20\% of lenses are produced by spiral galaxies
(e.g.\ Fukugita \&~Turner 1991),
so hundreds of lensed systems must be found to provide a
useful sample of spiral galaxy lenses.

Fortunately, there is no shortage of undiscovered lenses.  
Statistical estimates suggest that there are $0.5$ lenses per
square degree for a quasar sample complete to about $V$=24 (Kochanek 1996a),
or for radio sources with 5~GHz radio flux of $>$10~mJy (Kochanek 1996b). 
Equivalently, 1--10 in 1000 sources with redshifts above unity will be
gravitationally lensed depending on the type of source and its flux.
The problem is finding them in an {\it efficient} manner.

The first surveys for lenses focused on bright quasars (see Blandford
\&~Narayan 1992), with a discovery rate of roughly 1 per 100--200
targets.  The high rate is due to the effects of magnification bias
(Turner 1990), which is important for all lens samples but critical
for surveys of bright quasars.  
This strategy is limited by the relative rarity of bright quasars, the
difficulty of obtaining large numbers of high resolution optical
images, and contamination from stars.

Another successful strategy was to look for lensed radio sources.
Extra-galactic radio sources are numerous and distant, 
and confusing low-redshift sources are rare.
The three largest radio searches, MG-VLA (Burke\etal\ 1993),
JVAS (Patnaik\etal\ 1992; Patnaik 1993) and CLASS (Jackson\etal\ 1997),
have discovered almost half of the $\sim$60 known galaxy-mass lenses.
Their strategy was to image all radio sources to a fixed flux limit,
and then use the morphology of each radio image to identify the lenses.
Typically, one lensed radio source is discovered
for every 500 survey targets.  Radio searches also led to the
discovery of an important new class of lenses where the lensed images
of an extended background source merge to form an ``Einstein ring''
of emission around the lens galaxy (e.g.,~Hewitt\etal\ 1988).  Unlike a
compact source, an extended background source probes many lines of
sight which extend well into the halo of the lens, allowing much
better constraints on the dark-matter profile (e.g.,~Kochanek 1995).

In this paper, we present a much more efficient lens search,
which uses combined radio and optical surveys to find lensed radio lobes. 
The survey selection is outlined in \S2,
and follow-up observations are described in \S3.
In \S5, we discuss individual candidates for lensing, 
and in \S6, we summarize our conclusions.

\section{The FIRST-APM Lensed Lobe Search}

Most lensed radio sources are being missed by present searches.
Powerful radio galaxies are often flanked by two extended radio lobes,
whose radio spectrum is markedly steeper than that of the compact core
(e.g.\ FR-type objects; Fanaroff \&~Riley 1974). 
Due to their extended nature, the optical depth to lensing of such
lobes is significantly higher than for unresolved sources. 
Indeed, the typical lobe size of a few arcseconds implies that $\sim$70\%
of lensed radio sources should be lobes (Kochanek \&~Lawrence 1990).  
The JVAS and CLASS surveys select only flat-spectrum radio sources,
and thus lobes constitute only $\sim$20\% of their lenses.
The smaller MG-VLA search found more lensed lobes ($\sim$50\% of lenses),
but still fell short of the $\sim$70\% fraction. 
Many lensed lobes are simply not recognized, because the background source
itself often has complicated structure, and often lacks sharp features
which can produce a clear lensing signature.  Sensitive, high-resolution
radio maps are required to identify lensed structures 
($\sim$1\arcsec\ scale) in these cases. 

Lensed lobes can be found efficiently because they usually lie
{\it outside} their host galaxy.  Radio-galaxy hosts are centered on their
compact cores, so radio lobes should not have direct optical counterparts.
A galaxy which coincides with a radio lobe is almost certainly 
a chance alignment, 
and the relative distances of radio and optical sources
strongly favors the galaxy being a foreground lens
(Kochanek \&~Lawrence 1990).
Clear examples of this phenomenon were found in the MG-VLA search
(MG\,J1654+1346, Langston\etal\ 1989; MG\,J1549+3047, Leh\'ar\etal\ 1993).
Thus, an efficient way to find lenses is to identify 
radio lobes with superposed optical galaxies. 

A useful sample of radio lobes requires a large, low frequency,
radio survey with sufficient angular resolution to distinguish lobes from cores. 
The VLA\footnote{
     The Very Large Array (VLA) is an NRAO facility. 
     The NRAO is operated by Associated Universities, Inc.,
     under cooperative agreement with the National Science Foundation.
     }
FIRST Survey\footnote{see http://sundog.stsci.edu/}
(Becker, White, \&~Helfand 1995), 
has a flux sensitivity of 1 mJy ($6\sigma$) at 20\,cm,
$5\arcsec$ resolution (FWHM) and sub-arcsecond positional accuracy.  
This represents a factor of $\sim$50 improvement in both
sensitivity and resolution over previous large-area surveys.
At the time of our initial lens search, the northern FIRST sample 
covered $\sim$3000 square degrees 
(7$^h${}$<$RA$<$18$^h$ and 22$^\circ${}$<$DEC$<${}58$^\circ$),
and comprised $\sim$3$\times$10$^5$ radio components. 

A good sample of optical counterparts is provided by the 
APM\footnote{
     The Automatic Plate Measuring (APM) machine is a National
     Astronomy Facility run by the Institute of Astronomy in
     Cambridge, UK (see http://www.ast.cam.ac.uk/$\sim$apmcat).
     }
catalog (McMahon \&~Irwin 1992; Maddox\etal\ 1990), 
based on the Palomar Observatory Sky Survey plates (\mbox{POSS-I}). 
The \mbox{POSS-I} plates have a detection limit of $R${}$\sim$20, and
we conservatively estimate that 25--35\% of the known lens galaxies
(see Keeton\etal\ 1998) are bright enough to be included in the APM catalog.

Our initial search for lenses, of course, does not fully exploit
the potential of this method.  The shallow magnitude limit 
of the APM survey means that our lens sample is no longer mass-selected, 
given that only the optically brightest third will be included. 
Moreover, most of the FIRST lobes are much too faint to be followed up
at higher resolution, due to their steep spectra and extended structures. 
Nevertheless, we have searched the available FIRST survey down to its
detection limit, to get a sense for how many lensed lobes can be found this way. 

\subsection{Survey Selection}

For the FIRST-APM lensed lobe search, we combined these 
two surveys to find optical counterparts to radio lobes. 
We classified the FIRST sources by their radio morphology (see~Appendix),
and identified those components most likely to be lobes.
We then identified as lens candidates those radio lobes
with APM counterparts. 
The estimated mean redshift of FIRST sources 
is $\VEV{z}${}$\sim$1 (Cress \&~Kamionkowski 1998), 
and the measured redshift distribution of optical surveys such as 
the APM Galaxy Survey peaks at $\VEV{z}${}$\sim$0.12 (Baugh \&~Efstathiou 1993).
Thus it is likely that APM counterparts to FIRST radio lobes
are indeed in the foreground, and potentially lenses. 

To define a radio-lobe sample, it is necessary to first
``collapse'' the radio catalog, such that multiple components of a 
single radio source are associated with each other. 
The collapse algorithm is described fully in the Appendix.
We ran the collapse routine using a $60\arcsec$ search radius,
a maximum component separation of $90\arcsec$ for triples and multiples
(defined as sources with more than three components),
and a 10\% random association probability cutoff, $P<0.1$.  
For ease of detection and phase calibration in follow-up radio observations,
we required all component fluxes to exceed 3\,mJy, and at least one
component to exceed 10\,mJy. 
To enhance selection of true, FR-type double-lobed objects,
we required a symmetry factor of $S>0.5$ for triples
(where $S$ is defined as the ratio of the shorter
core-to-lobe distance to the longer), 
and a configuration angle of $\theta<30^\circ$ for triples and multiples
(see Figure~10 in Appendix). 
While these requirements will necessarily eliminate some true 
FR-type radio galaxies, this selection favors easily-identified
radio galaxies with the most typical component configuration. 
The output of the routine is a list,
containing positional, flux, and morphological data,
of collapsed radio sources containing two, three, or more components, 
each of which is identified as either a ``core'' or a ``lobe.''
Due to the wide array of radio-source morphologies (core-jet, double-lobed,
cometary, diffuse, etc.) and the limited resolution of FIRST,
there is no simple, obvious manner in which one can correctly
and unambiguously collapse all sources. The parameters we have chosen 
yield a reasonable balance between ensuring
an accurate correspondence between collapsed and real sources, 
and are in good agreement with the conclusions 
of ``by-eye'' inspections (D.~Helfand, private communication).

The APM-counterpart selection was as follows.
For every collapsed FIRST source, we searched for optical counterparts to 
each lobe component, as well as to the core component (or to the 
source centroid, if no radio core was identified).
Optical objects were taken to be coincident with a radio
component if they fell inside an
ellipse defined by the deconvolved FWHM major and minor axes of the 
radio component, as listed in the FIRST catalog. For cases
where the catalogued major and/or minor axis was less than $2\arcsec$, we
adopted an effective FWHM value of $2\arcsec$ for that axis. 
For sources without a radio core, optical objects that fell within $\ell/4$
from the average component position were considered core counterparts
(where $\ell$ is the maximum component separation).
Those sources with at least one optical counterpart to a radio lobe,
which was not also a counterpart to the radio core, became candidates for lensing.
We eliminated candidates in which a single optical counterpart matched
to both lobe components of a double radio source, as these usually corresponded
to identifications of the host galaxy;
for the same reason, we also eliminated cases where a double or
multiple radio source was found to have only one optical match, 
corresponding to a compact FIRST component. 
To ensure sufficient contrast for lensing signatures,
we eliminated all lensed radio lobes larger than $10\arcsec$ (FWHM),
and further eliminated all optical lens candidates whose lobe counterparts
fell farther than $2''$ from the fitted peak of the radio lobe.
Each lobe optical counterpart was assigned a letter classification (Table~1)
to indicate its likelihood of being a lens (lenses are likely to be red and extended), 
based on its APM color and morphology.

There were 1037 candidates in our sample,
each of which has an APM object within 2$''$ 
of a presumed FIRST radio lobe's peak. 
The selection methods applied to the radio lobe and optical counterpart samples
are summarized in Table~2.  
With about one lens candidate
for every $\sim$200 independent FIRST sources,
our candidate selection rate using our chosen criteria is roughly twice 
the discovery rate for earlier radio lens searches, 
which suggests that many of our candidates are not lensed.
Only about 14\% of the candidates have optical core detections,
compared to the much higher 40\% \mbox{POSS-I} identification rate
for MG-VLA lobed radio sources (Leh\'ar 1991),
partly because of the higher MG-VLA flux cutoff (60\,mJy at 6\,cm),
but also because many lensed-lobe candidates 
are themselves misidentified radio cores. 
The lensed-lobe candidate and FIRST optical counterpart distributions are similar,
and both have more red, extended objects than the APM catalog in general
(see Table~1), probably
because both radio samples preferentially select galaxies. 
There are a large number of blue counterparts to presumed radio lobes,
and probably many of these are quasar counterparts to radio cores.
However, 62\% of these blue counterparts have uncertain optical 
identifications, because they are close to the blue \mbox{POSS-I}
detection limit and are absent in the red catalog.

\subsection{The VLA Sample}

Because of the limited angular resolution of FIRST ($\sim$5$''$ FWHM),
higher-resolution radio observations are needed
to confirm candidates as lensed.
Most radio lobes extend over only a few arcseconds, so FIRST cannot
always determine unambiguously which radio components associated
with a source are truly radio lobes, and which are compact cores. 
Moreover, the typical image separation, 
$\Delta\Theta$, for galaxy-mass lenses is only $\sim$1$''$, 
so the effects of lensing cannot be seen directly in the FIRST images. 
The VLA can achieve sufficient angular resolution 
to distinguish lobes and resolve gravitational lensing, 
but only by observing at shorter wavelengths 
(A-array resolution $\sim$0$\farcs$4 at 6\,cm).  

Since radio lobes generally have very steep radio spectra, 
the reduced source flux at shorter wavelengths
drastically reduces the number of lens candidates 
accessible to short, high-resolution VLA observations. 
The lensed radio lobe in MG\,J1549+3047 (Leh\'ar\etal\ 1993) 
is typical of what we expect to find in our lensed lobe search, 
in that only the fainter parts of the lobe are multiply imaged
by an off-center lens galaxy.
The lensed lobe has a FIRST flux density of 877\,mJy,
but at 6\,cm the total flux density is 200\,mJy, 
with the hot-spots peaking at only 16\,mJy per beam. 
The mean surface brightness of the lensed ring is much fainter
(typically $\sim$2\,mJy/beam), but can be detected
with a signal-to-noise ratio of 3
in a 1.5\,minute VLA observation. 
If our lensed lobes have similar properties, lensing signatures 
should be detectable in 30\,minute observations
for FIRST lobes exceeding 200\,mJy. 
A lensed lobe like that in MG\,J1549+3047, 
but with a FIRST flux of 100\,mJy, would require
a 2\,hr VLA 6\,cm observation for reliable ring detection. 
Much shorter observations, however, would suffice to rule out
cases where we had misidentified a compact core as a lobe. 

With these considerations, we selected a small sample
of lensed-lobe candidates for VLA confirmation. 
There were 78 lensed-lobe candidates with FIRST flux $>100$\,mJy.
The limited VLA schedule forced us to choose a subsample
of these targets, and this was done by visual inspection.
By comparing the FIRST maps to \mbox{POSS-I} fields from the
Digitized Sky Survey (DSS\footnote{
    Based on photographic data of the National Geographic Society -- 
    Palomar Geographic Society to the California Institute of Technology. 
    The plates were processed into the present compressed digital form 
    with their permission. The Digitized Sky Survey was produced at the 
    Space Telescope Science Institute under US Government grant NAG W-2166.}),
two of us (AB and JL) visually ranked each of the 78 candidates
by likelihood of lensing, based on optical colors, radio morphology, 
alignment, and so on. We compared notes on our rankings and
eliminated those candidates that seemed the least likely to be lensed
(e.g., candidates which had a single APM counterpart corresponding
to a fairly compact component between two others).

Our final VLA sample of 33 candidates is listed in Table~3,
and, as a visual guide, Figure~1 presents FIRST cutout maps,
with optical fields from the DSS.  
Since the DSS astrometry can be in error by a few arcseconds, 
we registered the DSS images to the APM catalog, whose absolute
positions deviate by $<$0$\farcs5$ for small objects.
This was done by selecting $\sim$20 APM objects in each DSS field
and finding the least-squares linear coordinate transformation
on the DSS images
(internal residuals were typically less than $0\farcs5$).
The FIRST maps were superposed using the VLA phase calibrator
astrometry which routinely yields $\sim$0$\farcs2$ positional accuracy.

\section{Observations}

Most of the FIRST/APM lensed-lobe candidates are expected to 
be radio cores which had been misidentified as lobes,
or genuine radio lobes with stellar or nonexistent counterparts
(APM/POSS artifacts).  To eliminate such cases, and to 
establish remaining gravitational lens candidates, we obtained 
higher quality radio and optical observations.  
We used archival data where available, but the bulk
of our targets required new observations.  
The observations we used are summarized in Table~4, 
and will be referred to by the labels therein.

We obtained VLA observations for the accessible sample (Table~3). 
Nine of our targets had appropriate archival observations,
and we observed the other 24 using the A-array at 6\,cm (r98b, see Table~4).
The total allocation of 8~hours was divided between the targets
based on the lobe flux (see Table~3).
Calibrators were observed every 
$\sim$15\,min to ensure proper phase calibration,
and the flux densities were scaled to 3C286
(with an assumed flux density of 7.43\,Jy).
Calibration and mapping followed standard AIPS\footnote{
   AIPS (Astronomical Image Processing System) is distributed by 
   the National Radio Astronomy Observatory, which is a facility
   of the National Science Foundation operated under cooperative 
   agreement by Associated Universities, Inc.} 
procedures, with a single iteration of self-calibration, and a circular 
convolving beam of $0\farcs7$ (FWHM), to bring out extended structures.
The resulting maps have off-source rms noise levels of $\sim$0.1\,mJy/beam,
close to the thermal noise limit, except for the brightest targets. 
As expected, most of our radio lobe candidates turned out to be 
misidentified radio cores, and sometimes presumed lobe
components could be resolved into separate radio sources.

Three of the most promising VLA targets
were observed at 18\,cm using MERLIN (r98a, see Table~4). 
MERLIN provides sub-arcsecond resolution at lower frequencies, 
where the steep-spectrum radio lobes are brightest,
but it is not very sensitive to faint sources. 
J0823+3906, J0828+3552, and J1622+3531 were each
observed eight times in 30\,minute scans at intervals
of 2--6~hours, to maximize baseline sampling.
Standard MERLIN phase calibrators were also observed, 
and the flux densities and polarizations were calibrated 
using 3C286 and OQ208, respectively.
The data were reduced using the standard MERLIN pipeline, 
with several iterations of phase self-calibration. 
The resulting maps have off-source rms noise levels of $\sim$0.2\,mJy/beam,
and an angular resolution of $\sim$0$\farcs$25 (FWHM).
MERLIN produced an excellent map of J1622+3531, but in most cases,
faint radio lobe structures could be seen.

We obtained optical CCD images for the 12 most promising targets.
The objective was to confirm whether candidate lenses are galaxies,
and to seek optical counterparts to the radio cores.
Observations were acquired when the opportunity arose, 
so a variety of telescopes and filters were used (see Table~4).
The data quality was variable, but the integration times 
were chosen to permit detection of an $R${}$\sim$24\,mag galaxy. 
Photometric calibration was achieved, in most cases, by acquiring
Landolt (1983) standard star observations during each run,
but in a few cases we were forced calibrate either to spectra
or to the APM photometry.  In all cases, photometric accuracy is
no better than 0.1\,mag on an absolute scale, 
although relative photometry in each image will be considerably better. 
The images were reduced using standard IRAF\footnote{
   IRAF is distributed by the National Optical Astronomy
   Observatories, which are operated by the Association of Universities for
   Research in Astronomy, Inc., under cooperative agreement with the National
   Science Foundation.}
procedures. 
Over half of the observed candidates could be eliminated as
either stars or spurious APM objects, and radio cores were detected
for all of the remaining lens candidates (see Figure~2). 
Two of our targets (J0823+3906 and J1605+5439) were observed through
several filters, though time considerations did not permit this for most.

Four of our targets (J0823+3906, J0849+2448, J0958+2947, J1622+3531)
were also observed spectroscopically, again using a variety of telescopes
as the opportunity arose (see Table~4).  The spectra were reduced using
standard procedures, and were calibrated by observing flux density standards. 
All but one of the observations yielded unambiguous redshifts for the targets.

\section{Individual Targets}

Here, we discuss in detail the twelve candidates which could
not be rejected immediately as misidentified radio cores 
and received CCD observations (see Table~3 and Figure~2). 
We will refer to observations using the labels in Table~4,
and optical photometry of the counterparts will be compared to 
passively evolving early- and late-type galaxies at various redshifts
(for details, see \S5 of Leh\'ar\etal\ 2000). 

J1549+3047 is the known lens MG\,J1549+3047 (Leh\'ar\etal\ 1993).
This target was selected by our search method,
and serves as a clear example of a successful detection.
The APM finds a bright counterpart to the western FIRST component.
Here, we present a 45\,s excerpt of the 6\,cm VLA data
from Leh\'ar\etal\ (1993), which gives a lensed ring detection
which is comparable to what our other observations should yield.
The CCD image clearly detects both the lens galaxy and the core counterpart,
and the measured lens redshift ($z$=0.111, Leh\'ar\etal\ 1993)
is much less than the typical radio source redshift. 
Moreover, corresponding parts of the lensed ring have matching spectral 
index and fractional polarization (Leh\'ar\etal\ 1993). 

J0816+5003 is lensed.
There are three optical counterparts, one on each of the 
FIRST components, including the central core component (Figure~2).
The $R${}$<$16 southern lobe counterpart is stellar, 
and is too far away from the radio lobe to be of interest to us.  
However, the northern lobe counterpart, Galaxy~1, 
lies directly over the northern radio component,
which is broken into three VLA components (N1--N3, Figure~3).
The unusual structure of the Northern FIRST component is
confirmed in archival multifrequency VLA observations 
(r96: 2\,min at 20\,cm, 5\,min at 6\,cm, 8\,min at 3.6\,cm). 
We present a lens model in which N1 and N2 are 
lensed images of a northern lobe hot-spot, 
and N3 is the northern lobe's unlensed terminal ridge.
The N1--N2 separation of 1$\farcs$9 corresponds to 
an isothermal $z${}$\sim$0.4 lens galaxy with 
a velocity dispersion of $\sim$250$\kms$. 
The lens model has been matched in position and shape to Galaxy~1 
(deconvolved FWHM of $\sim$1$\farcs$8, axial ratio $\approx$0.9, 
and PA=--64$^{\circ}$). 
An alternative view is that N1--N3 compose a separate radio source,
with its core at N2 corresponding to Galaxy~1. 
While we cannot completely exclude this possibility,
we favor a lens interpretation for the following reasons:
(1)~The clear lobe-like appearance of the southern FIRST component 
leads one to expect a northern radio lobe near Galaxy~1.
(2)~Component N1 has a linear morphology which is very unusual for a
radio lobe, yet fully expected in our lens model. 
(3)~N2, which would be the core of a separate FR-type source,
does not coincide with Galaxy~1's center (N2 is $0\farcs4$ to the North),
when the radio and optical images are aligned using the core components.
N2 could be an off-set radio jet, but such jets are less common
than coincident cores.  Furthermore, most radio jets appear on the
same side as the brighter and more polarized radio lobe
(see~Saikia \&~Salter 1988),
unlike the situation for N1--N3.
This arrangement, however, is naturally explained by our lens model.
(4)~N1 and N2 have identical radio spectra which are 
more similar to the other lobe hot-spots than to the flatter
radio Core's (Figure~4).  
Radio lobe hot-spots usually have steeper spectra than radio cores, 
and their spectra often differ within a single radio galaxy. 
Similar spectra are expected, however, 
for gravitational lensing, which is achromatic. 
(5)~N1 and N2 have the same polarized emission properties,
both in fraction and orientation, which are very different from N3. 
Normally, there is no correlation between the core and lobe
polarizations (see~Saikia \&~Salter 1988),
but lensed image polarizations should match. 
(6)~The radio core counterpart is much fainter ($R$=21.7)
than Galaxy~1 ($R$=19.2), as expected if it is more distant,
and the optical magnitude of Galaxy~1 is consistent with an
$L_{*}$ early-type galaxy at intermediate redshift (0.3$<${}$z${}$<$0.5). 
So, the properties of N1--N3 are very unusual for an FR-type radio source,
but typical for a gravitational lens, with a morphology similar to
FSC~10214+4724 (Broadhurst \&~Leh\'ar 1995; Eisenhardt\etal\ 1996). 
Optical spectroscopy is planned to confirm this system as lensed,
and high-resolution radio observations are scheduled, 
to seek more arc-like lensed images, and to determine whether 
N1 is composed of three merged images.

J0823+3906 is a strong candidate, 
with a galaxy directly on a radio lobe, 
but there are no clear lensing signatures. 
The lobe counterpart (Galaxy~1) is clearly an extended galaxy, 
with a deconvolved size of $\sim$1$\farcs$2 (FWHM). 
If we align the optical image so that the optical core 
coincides exactly with the radio core, 
the section of the lobe under Galaxy~1 is too faint to allow 
multiple imaging to be detected in either the VLA or MERLIN maps.
Deeper VLA observations 
(14\,min at 20\,cm, and 1.7\,hr at 6\,cm, Figure~5) detect 
the flat-spectrum radio core ($\alpha>+0.5$), 
and find substantial emission near Galaxy~1,
but still show no lensing signatures 
in the steep-spectrum lobe ($\alpha${}$\sim$--1). 
Lensing would not occur, of course, 
if the radio source were closer to us than Galaxy~1.
Palomar~5\,m spectroscopy (Figure~6) could not detect the core,
and there were no clear features in the spectrum of Galaxy~1. 
However, several dips in the continuum suggest $z${}$\sim$0.3 as a possibility.
V~and I~band images of J0823+3906 (from o00b),
yield $I$=19.2 and $V$--$I$=1.6 for Galaxy~1. 
This is consistent with an $L<L_{*}$ early-type galaxy at $z${}$\sim$0.3
or an $L${}$\sim${}$L_{*}$ spiral at $z${}$\sim$0.5\,. 
Either way, Galaxy~1 is not likely to be farther than $z${}$\sim$0.6\,. 
The distance to the radio source is harder to determine, however,
although it is much fainter ($I$=21.8) than Galaxy~1. 
To eliminate lensing for J0823+3906,
deeper optical spectroscopy is crucial. 

J0849+2448 is not a gravitational lens.
There are optical objects close to both radio components,
but their $12\farcs7$ angular separation is greater than
the radio component separation, and neither of them falls
directly on a radio component.  Moreover, both are optically 
compact, and optical spectra (s98) show them to be local
M~(North) and G~(South) stars.

J0958+2947 is not lensed, 
but includes two quasars at very different redshifts. 
This source was first noted in the MG-VLA search 
(MG0959+3047, Leh\'ar 1991). 
The VLA map resolves the FIRST components into two separate quasars,
each with cores and radio lobes, and there are compact, blue optical
counterparts to each core.  
The optical spectra of the cores (Figure~8) show emission lines
corresponding to quasars at high redshift 
($z_{north}$=2.064 and $z_{south}$=2.744).
J0958+2947 does not show any clear signs of lensing,
but the 20\,cm VLA map (Figure~7) and the 6\,cm map both show 
a small arc-like extension, N5, about 2$\arcsec$\ northeast of N1.
This could either be an unusual structure connected with this quasar, 
or it could be part of the of the northern lobe associated with S1,
which is otherwise not seen.  
The spectral index ($\alpha${}$<$--0.5) and polarization ($\sim$5\%) 
of N5 are typical for a lobe hot-spot. 
However, the high redshift of the Northern QSO makes it 
an inefficient lens for a source at $z$=2.744, 
so $\Delta\Theta${}$\lax$0$\farcs$4 for an $L_{*}$ galaxy.
Thus, even if N5 is part of the southern radio source, 
we would not expect to see multiple imaging. 

J1320+4639, J1342+3507 (4C+35.30), J1357+4807, and J1425+3027 (MG\,J1425+3027)
are not gravitationally lensed.  In each case,
deeper optical observations failed to detect any lobe counterpart, 
so these are likely to be spurious APM detections. 

J1605+5439 is not lensed. 
The southern lobe counterpart has a stellar profile, 
and there is no evidence of lensing in the radio lobe. 
Moreover, the photometry (from o00b) 
of the lobe counterpart ($I$=16.8, $V$--$I$=3.2) 
is inconsistent with a normal galaxy at any redshift. 
The core counterpart ($I$=20.3, $V$--$I$=1.6), however, 
is consistent with a galaxy at $z>0.3$.

J1622+3531 has nearby galaxy in front of a distant lobe,
but our observations failed to detect lensing signatures.
The VLA and MERLIN (Figure~9) maps show the northern lobe 
to be broken into two steep-spectrum ($\alpha${}$\sim$--1)
subcomponents, N1 and N2,
and there are optical counterparts to both the northern lobe
and the inverted spectrum radio core ($\alpha$=+0.5$\pm$0.1).
Galaxy~1, at $z$=0.32, is extended (deconvolved FWHM $\sim$0$\farcs$9)
and at lower redshift than the Quasar, at $z$=1.47 (Figure~10).
We can roughly reproduce the configuration of N1 and N2,
using a simple lens model (Figure~9), 
and their spectral indices are similar
($\alpha_{N1}${}$\approx$--0.4, $\alpha_{N2}${}$\approx$--0.6),
as expected for lensing. 
However, the fractional polarizations (Figure~9) of N1 and N2
are inconsistent with lensing.
Moreover, the lens must lie between N1 and N2,
where no optical counterpart is found.
We attempted to model the radio structure using models with two 
masses at the positions of Galaxy~1 and Galaxy~2, but these invariably 
produced unobserved bright counter-images near Galaxy~1. 
Thus, we conclude that the morphology of the northern lobe 
is probably due to a bent radio jet (e.g.\ Lonsdale \&~Barthel 1986).
Galaxy~1 {\it must be} lensing the more distant radio lobe, 
but it is probably just distorting the intrinsic shape. 
At $z$=0.32, Galaxy~1  has an expected image separation 
of $\sim$0$\farcs$7, so we might find a faint lensed image 
of the northern hot-spot near Galaxy~1 in deeper radio observations. 

J1641+5115 is not a gravitational lens.
There are three APM counterparts, one each for the two FIRST components,
and one between them.  The VLA map shows the western radio component to be
the core, and the eastern counterpart falls 2$''$ away from the eastern radio source.

\section{Conclusions}

Our initial search for lensed radio lobes 
using the FIRST survey and the APM catalog 
has produced two lenses: J1549+3047 is 
the known lens MG\,J1549+3047 (Leh\'ar\etal\ 1993);
and J0816+5003 is a new lensed radio lobe system
that requires only spectroscopic confirmation. 
Two, more tentative, lens candidates (J0823+3906, J1622+3531), 
have galaxies superposed on radio lobes,
but show no evidence of multiple imaging.
One of our targets (J0958+2947) is a pair of quasars in close projection.

Our search method is highly efficient, even assuming that all 
of our tentative candidates are not lensed. 
The FIRST/APM search found 2 lensed systems,
using only 33 VLA observations. 
Thus 6\% of our VLA targets are lensed,
compared to a success rate of only 0.2\% for other radio lens searches,
and $<$1\% for bright quasar searches.
While these efficiency estimates do not take into account the effort
spent on follow-up optical and radio observations, it is clear that
combining optical and radio survey information can drastically reduce
the required number of targets for detailed observations. 

There are many more lensed radio lobes to be found in this way. 
Since our initial search, the FIRST survey has trebled in size,
covering $\sim$80\% of the planned $\sim$10,000 square degrees. 
Thus, we should discover 5--10 lensed lobes 
using the FIRST/APM comparison in the very near future.  
 
The sample can be expanded and improved with fainter optical catalogs.
The \mbox{POSS-I} plates detect only $\sim$30\% of known lens galaxies, 
so the resulting lens sample is far from mass-selected. 
However, almost all the known gravitational lenses are brighter than $R$=23,
and should be accessible to optical surveys (see Keeton\etal\ 1998).
We are currently extending our search using the \mbox{POSS-II} survey
(Djorgovski\etal\ 1998), which should find
{\it at least} $\sim$50\% more lens galaxies than the APM,
leading to $\sim$50\% completeness, and 7--15 expected lensed lobes. 
Moreover, two thirds of the APM lobe counterparts
turned out to be stellar or spurious in CCD images.
This suggests that selection based on a higher quality optical survey,
such as the Sloan Digital Sky Survey (Gunn \&~Knapp 1992),
could treble the search efficiency.

Even greater gains are possible as radio telescopes improve. 
We had to reject 92\% of our FIRST/APM candidates because the FIRST 
components were too faint for easy VLA mapping.  
Thus, as the VLA upgrades to larger bandwidth, 
and once higher resolution 20\,cm observations are possible 
using the VLA-Pie~Town link or the extended A+~array, from scores to 
hundreds of lensed lobes could be found in the Northern sky.

\section{Appendix: Collapsing the FIRST catalog}

Many radio sources exhibit complex morphology
(e.g., double-lobed, FR-type objects), and often include
several components in the FIRST catalog.  Thus, it is useful to first
``collapse'' the FIRST catalog into component groups for each source.
The collapsing procedure we employ here is based on the
circular overdensity algorithm of Buchalter (1999),
which has been used to systematically identify
and characterize large numbers of multi-component radio
sources in the FIRST Survey (e.g.~Blanton\etal\ 1997; Buchalter\etal\ 1998).

First, we morphologically classify individual FIRST components
and collect them into associated groupings.
Components classified as side-lobes are removed from the radio
catalog and remaining components are classified as ``extended'' or
``compact'', depending on whether the major axis of the fitted
Gaussian profile exceeds 2\arcsec. 
For compact (extended) components, the radio flux is taken to be the
listed peak (integrated) flux (see Becker, White, \&~Helfand, 1995).
Starting at the coordinates of the first catalogued component in a source, 
the program looks for other components within a fixed search radius, $R$.
If another component is found, the average position is determined,
and the search continues from the new average position. 
The process is repeated until no additional components are found 
within a distance $R$ of the last search center.

For each component grouping,
an estimated probability of random association is calculated
to determine whether the grouping qualifies as a single radio source.
Denoting the number of sources found in a given iteration by $N$, 
the largest separation between
the original source and any other by $d_1$, and the flux of the faintest
source in a given iteration by $F_{\rm min}$,    
the program calculates the probability of obtaining
$N-1$ sources all with fluxes $\ge F_{\rm min}$ and
within a distance $d_1$ of the original source, based on counting
statistics and on the measured two-point correlation function
of the FIRST Survey\footnote{
    The two-point correlation function, $\xi_2$, gives only the
    excess probability of finding {\em pairs} separated by a
    given distance. Inasmuch as the observed clustering is
    weak, i.e., $\xi_2 \ll 1$, we may conservatively take
    $\xi_2$ to be an upper limit for the excess probability of
    finding configurations of triples, quadruples, and so on,
    since higher-order $n$-point correlation functions,
    $\xi_n$, scale roughly as $\xi_n \propto {\xi_2}^{n-1}$ in
    the quasi-linear regime (Fry 1984). According to Cress\etal
    (1996) the measured two-point correlation function for
    FIRST, once the fragmentation of multicomponent sources
    (such as FR-type objects) is accounted for, extrapolates to
    a value of unity only at a scale of about 20$\arcsec$, so
    that using $\xi_2$ as an upper limit to the higher-order
    $n$-point correlation functions should be valid at all but
    the smallest scales we consider. Even if two components are
    separated by only the survey resolution limit of 
    $\sim$5$\arcsec$, and the two-point function exhibited no turnover
    at such small scales, we would only find $\xi_2${}$\approx$4,
    thus perhaps slightly overestimating the number of
    collapsed sources at the very smallest scales. }
(Cress\etal\ 1996).
If the observed configuration has a small probability, $P$,
of random association, compared to a chosen significance cutoff, 
the components are considered to be associated with a single source. 
Once a component is identified with a source, it is no longer
considered by the collapse routine in any subsequent searches.
If, however, the random association probability exceeds the cutoff,
such that no association is found,
the program initiates a new search centered at the position of the
next catalogued FIRST component.  The individual components in the
``failed'' search attempt are still considered as candidates
for collapse in later search iterations. 

Each source is then characterized by its component properties and
a set of parameters which can vary with the number of components $N$. 
For all sources, the average position and aggregate radio flux
of all components are tabulated, as well as the flux of
the brightest component. For pairs ($N$=2) the morphology
is characterized as ``both extended'', ``both compact'', or 
``one compact/one extended'' components. 
For the case of ``both extended'' components,
the difference in position angle between the two components
is calculated as a measure of the alignment. 

For triples ($N$=3), we identify the ``core'' component as that which
is not an endpoint of the longest leg, $\ell$, of the three triangle
legs formed (see Figure~11).  
Denoting the three legs by
$\ell_{12}$, $\ell_{23}$, and $\ell_{31}$, such that
$\ell=\ell_{12} > \ell_{23} > \ell_{31}$,
we calculate a symmetry factor, $S$, given by $\ell_{31}/\ell_{23}$, 
as well as a configuration angle, $\theta$, defined geometrically by 
$\theta = \arccos{\frac{\ell_{12}}{\ell_{23}+\ell_{31}}}$. 
Thus, $\theta$ is the angle defined by going from one focus to the other
to the endpoint of the semi-minor axis of an ellipse whose foci 
($F_1$, $F_2$) are 
separated by a distance $\ell_{12}$ and whose major axis is equal to
$\ell_{23} + \ell_{31}$. 
This configuration angle is computationally simpler than 
the traditional bending angle for radio sources, with which it
should not be confused. 
Requiring that $\theta$ be less than some chosen value,
$\theta_c$, ensures that the ``core'' is located within the ellipse
defined by $\theta_c$.  The requirement that
$\ell_{12} > \ell_{23} > \ell_{31}$ implies that $\theta<60^\circ$.
In addition, the morphology of the ``core'' component and the
``lobe'' components are noted for triples.

Associations with more than three FIRST components are
termed ``multiples.'' For these sources, we find the largest separation,
$\ell$, between any two components, and take the configuration angle,
$\theta$, to be the maximum value obtained (in the manner described
above) from all triangles in which one side is given by $\ell$. 
Furthermore, if one or more components lie within a distance $\ell/4$ of the
average position, the program identifies as a ``core'' that component 
nearest to the center; otherwise, no core is identified.

\acknowledgments
{\bf Acknowledgements: }
We are grateful to Bob Becker, Emilio Falco, Mike Irwin, 
Don Schneider, and Ion Yadigaroglu
for providing optical spectra and CCD images. 
It is also a pleasure to thank 
David Helfand and George Djorgovski for many helpful comments.
JL and CSK are supported by the Smithsonian Institution.
This work was partially supported by NASA/HST grants
GO-7495 and GO-7887, and by NSF/AST-9802732.
%


\begin{figure}   
   \figurenum{1}
   \centerline{\psfig{figure=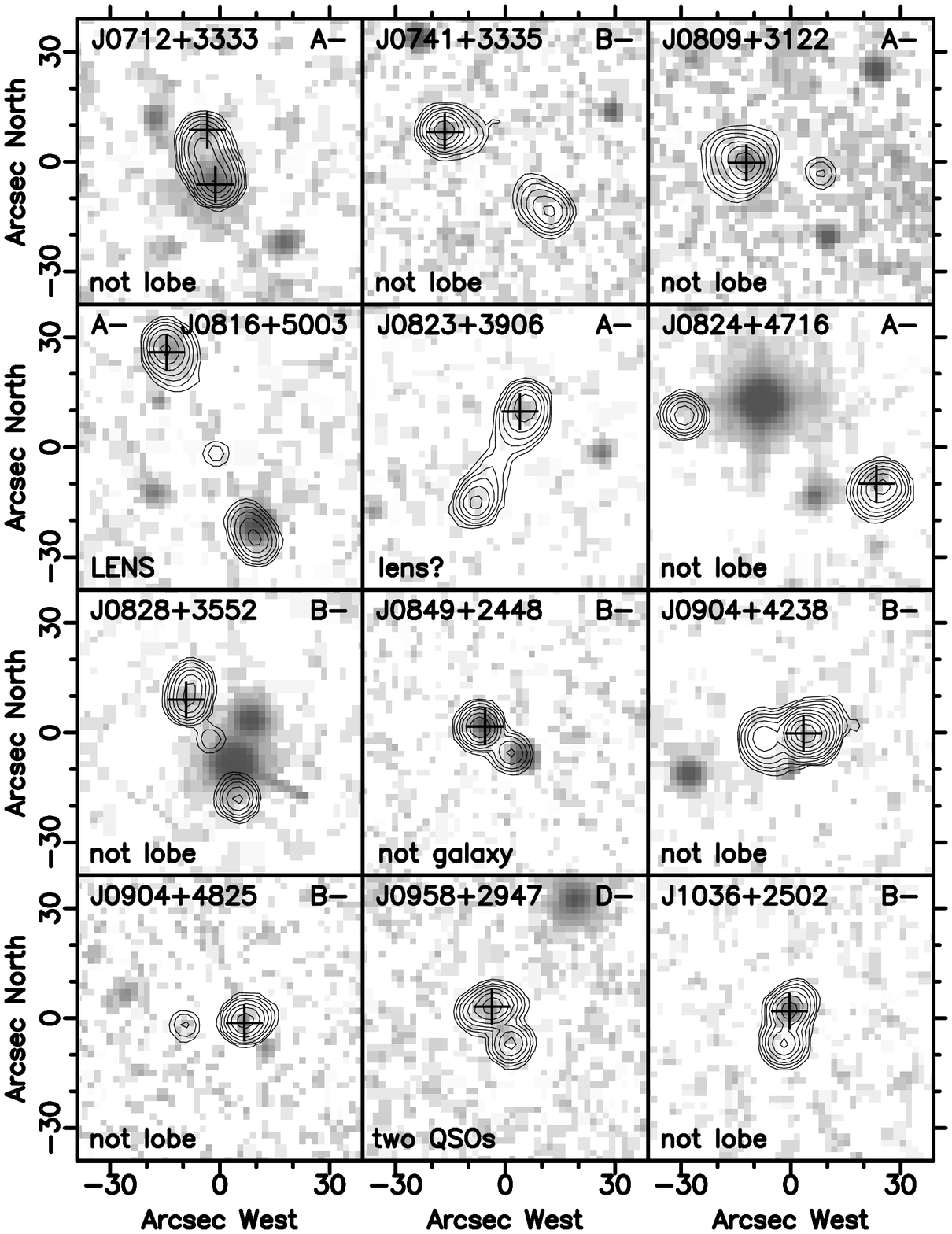,height=8in}}
   \caption{VLA sample of lensed-lobe candidates with the lobe counterpart's 
      optical class (Table~1), FIRST radio maps and DSS optical fields.
      Radio contours increase by doubling from 1~mJy/beam;
      and the grey~scale map shows the logarithm of DSS optical brightness.
      APM counterparts to FIRST components are marked by crosshairs.
      A few blue APM counterparts are not seen on the DSS. 
      The maps were aligned using APM astrometry, 
      and are centered on the positions given in Table~3.
      The outcome of our VLA and optical follow-up (see Table~3) 
      is also summarized.
   }
   \end{figure}
\begin{figure}   
   \figurenum{1}
   \centerline{\psfig{figure=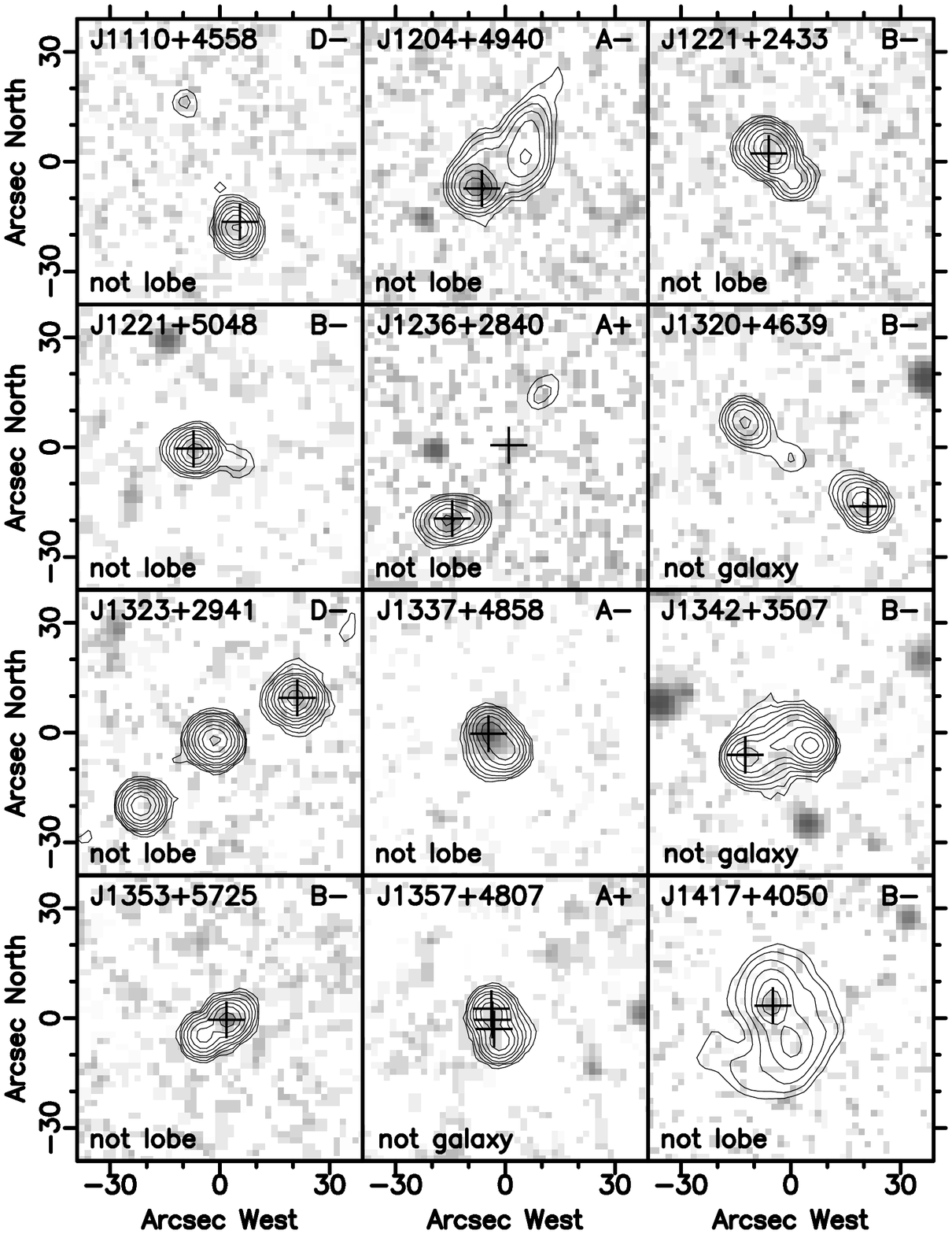,height=8in}}
   \caption{More FIRST/DSS maps of the VLA sample.}
   \end{figure}
\begin{figure}   
   \figurenum{1}
   \centerline{\psfig{figure=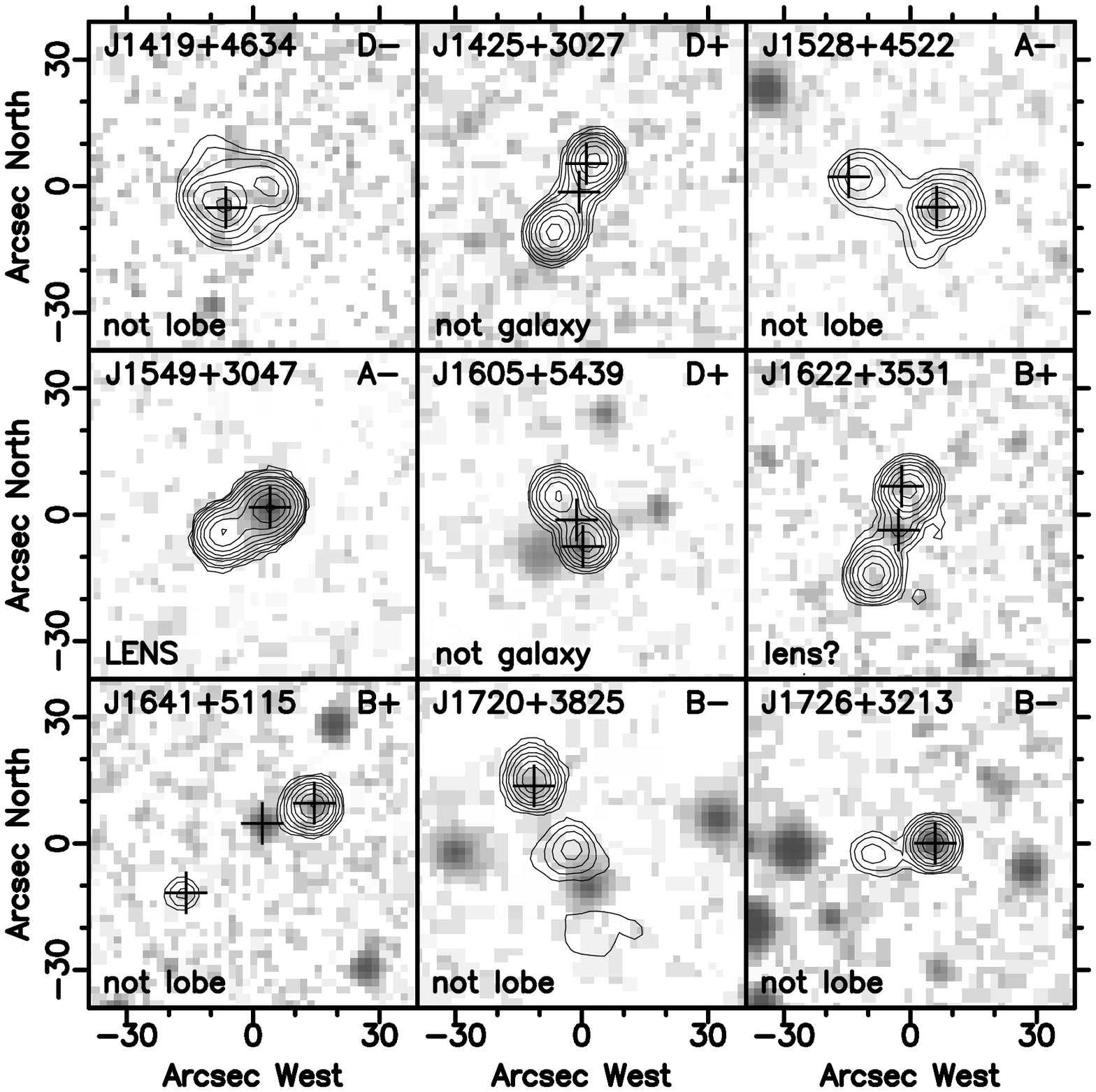,height=6in}}
   \caption{More FIRST/DSS maps of the VLA sample.}
   \end{figure}

\begin{figure}   
   \figurenum{2}
   \centerline{\psfig{figure=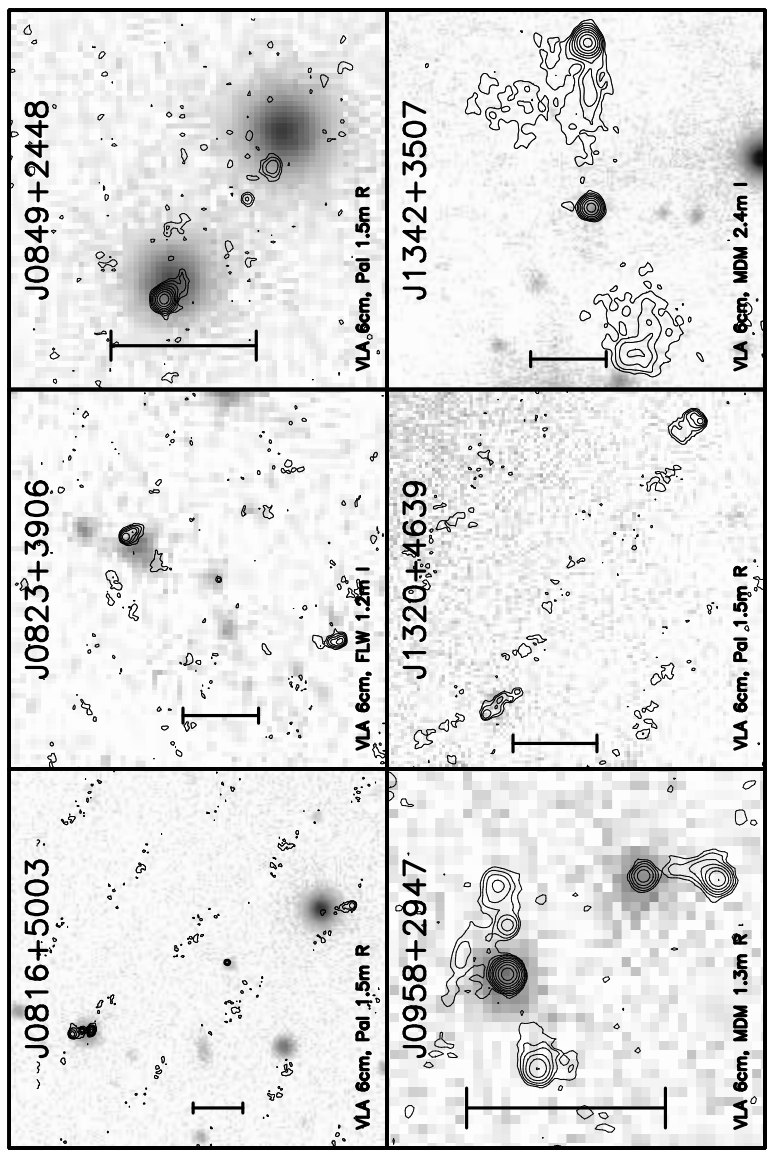,height=7in}}
   \caption{VLA and optical images of those candidates with CCD data (see Table~3).
      The radio contours increase by doubling from 
      twice the off-source map rms (Table~3) with a $0\farcs7$ beam size (FWHM).
      The CCD images, shown as logarithmic grey~scale, 
      were aligned using APM astrometry.
      The reference bars are 10$''$ long, with 2$''$ ends. 
      J0816+5003 is lensed, and J0823+3906 might be. 
      }
   \end{figure}
\begin{figure}   
   \figurenum{2}
   \centerline{\psfig{figure=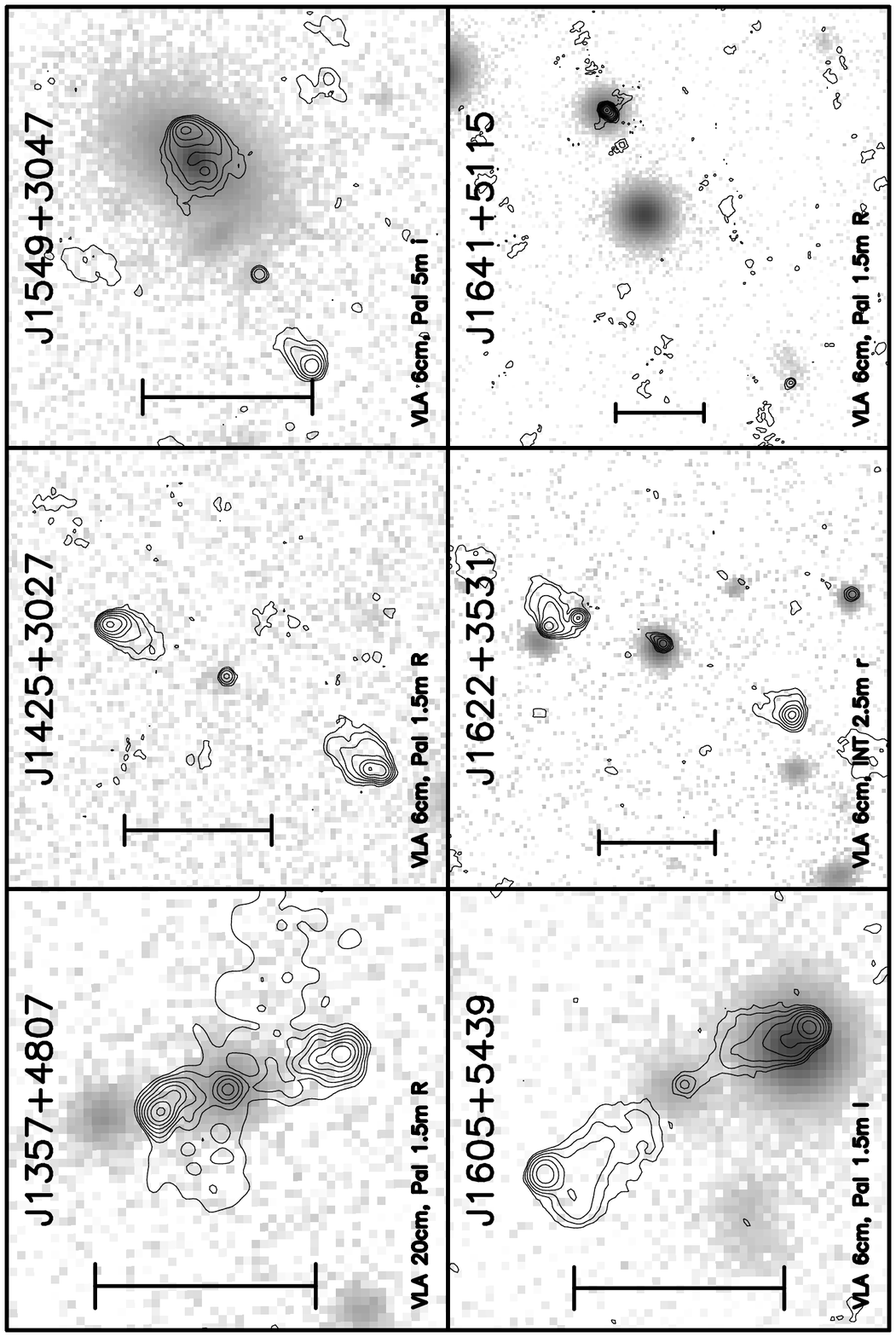,height=7in}}
   \caption{More VLA and optical images of CCD targets. 
   J1549+3047 is the known lens MG\,J1549+3047, 
   and J1622+3531 might also be lensed.}
   \end{figure}

\begin{figure}   
   \figurenum{3}
   \centerline{\psfig{figure=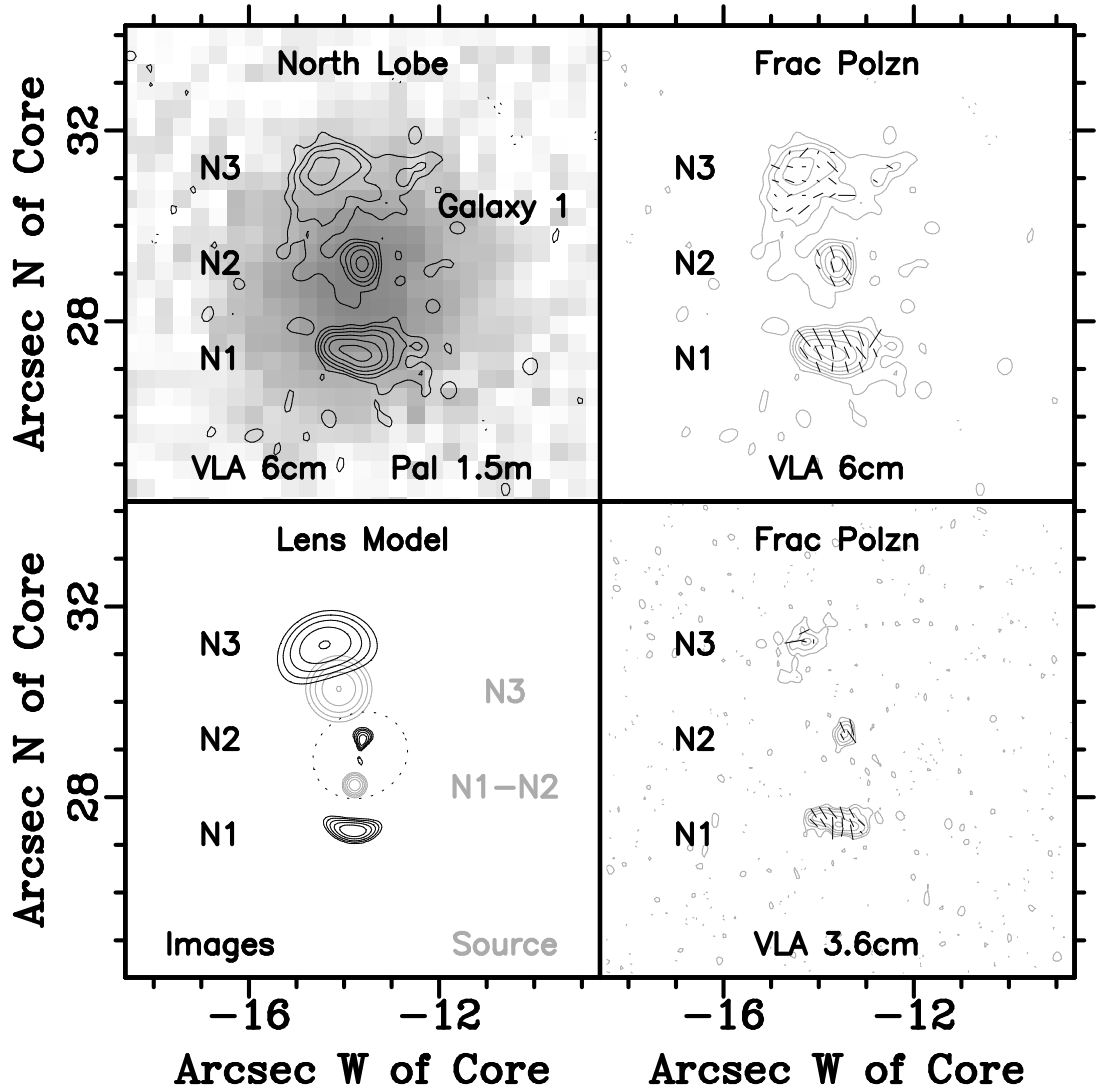,height=6.0in,angle=0}}
   \caption{
      High resolution VLA maps of the J0816+5003 
      northern lobe components (N1--N3).
      The 6\,cm maps (r98b) have a 0$\farcs35$ beam size,
      while the 3.6\,cm maps (r96) have a $0\farcs2$ beam (FWHM).
      The optical field (o00b, upper left) is shown in logarithmic greyscale,
      aligned to the radio using the core.
      Radio contours double from twice the 
      off-source rms noise of 0.055\,mJy\,beam$^{-1}$ (6\,cm),
      and 0.048\,mJy\,beam$^{-1}$ (3.6\,cm).
      The fractional polarization vectors for the northern component 
      (right panels) are parallel to the electric field, and scaled so that
      one spacing interval corresponds to 33\% polarization per beam.
      A singular isothermal ellipsoid model (lower left),
      matched to Galaxy~1, shows N1 and N2 as lensed images of a compact hot-spot.
      The unlensed source contours are shown in grey,
      while the lensed image contours and critical line (dotted) are in black.
      The 0$\farcs$2 positional discrepancy between the 3.6\,cm and 6\,cm maps
      is entirely accounted for by the coordinate system 
      precession between the two observations.
      }
   \end{figure}

\begin{figure}
   \figurenum{4}
   \centerline{\psfig{figure=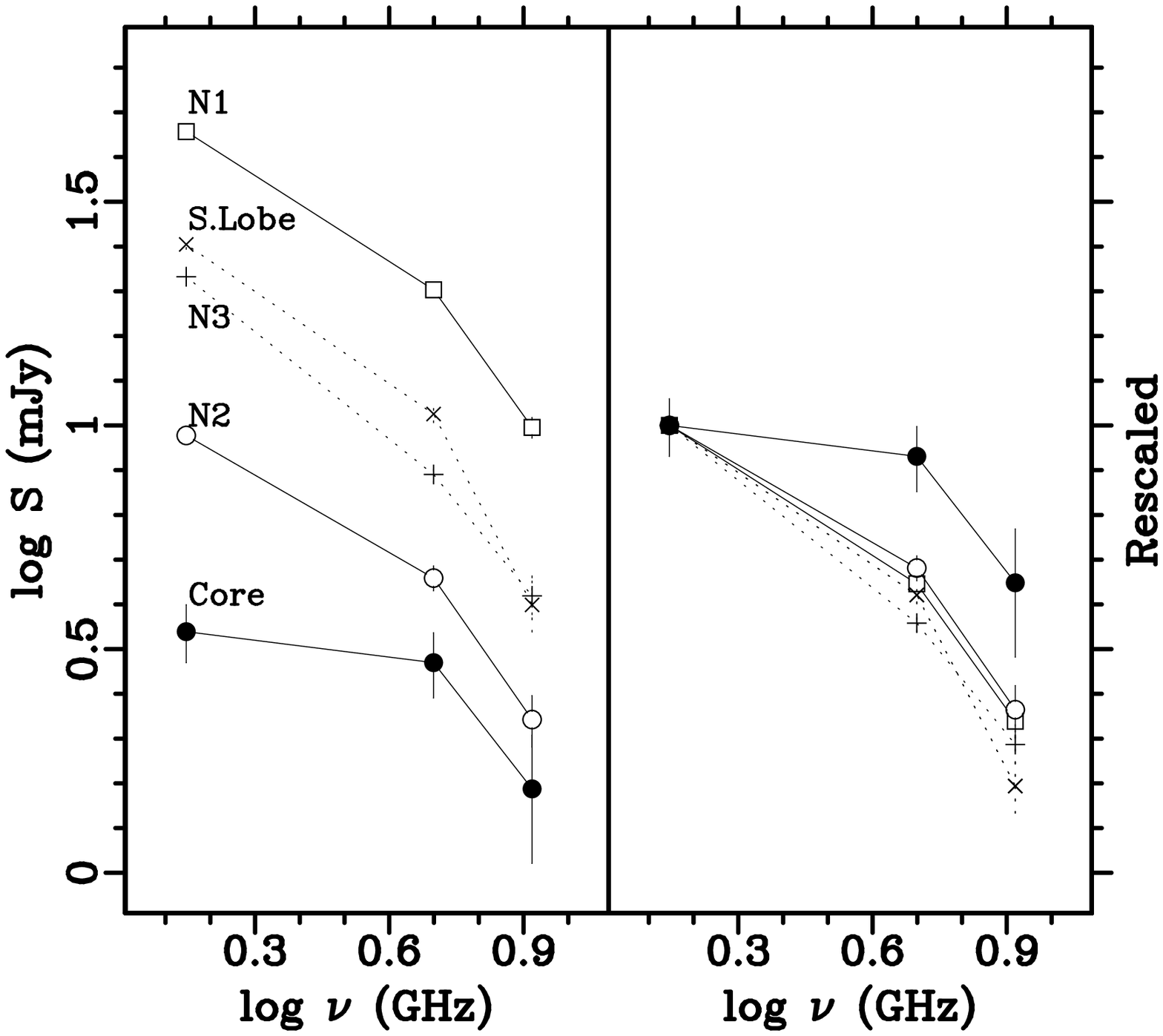,height=6.0in,angle=0}}
   \caption{
      Radio spectra of the J0816+5003 components,
      using archival multifrequency observations
      (r96: 2\,min at 20\,cm, 5\,min at 6\,cm, 8\,min at 3.6\,cm).
      Fluxes are presented for the northern lobe components,
      the radio Core, and the southern lobe's hot-spot.
      The left panel shows the actual component fluxes,
      and the right panel shows them normalized to the 20\,cm flux. 
      }
   \end{figure}

\begin{figure}   
   \figurenum{5}
   \centerline{\psfig{figure=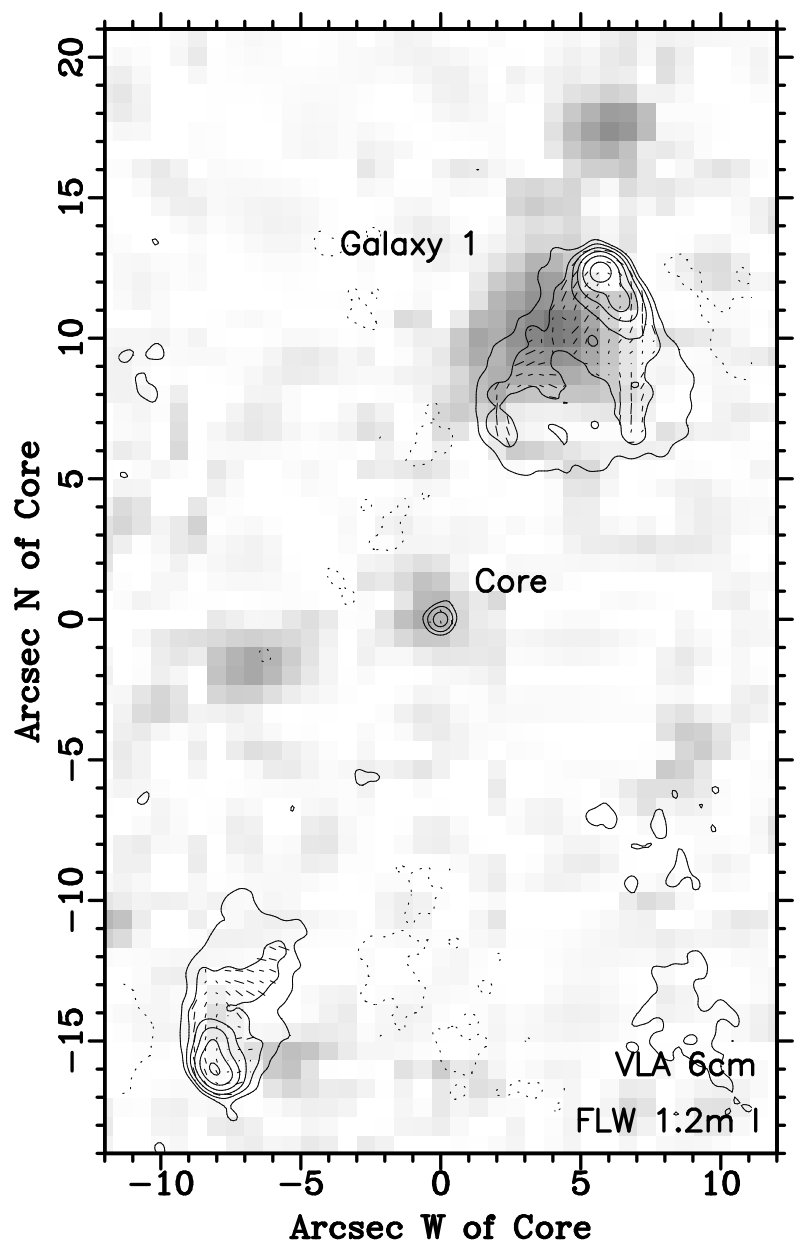,height=7.3in}}
   \caption{
      Deeper VLA map of J0823+3096 (r99) at 6\,cm (1.7\,hr).
      The optical field (o00a) is shown in logarithmic grey~scale, 
      aligned to the radio using the core. 
      The radio contours double from twice the off-source rms noise
      of 0.062\,mJy\,beam$^{-1}$, for a beam FWHM of $\sim$0$\farcs7$.
      The fractional polarization is shown as vectors parallel to the 
      electric field, scaled so that one spacing interval
      corresponds to 33\% polarization per beam.
      }
   \end{figure}

\begin{figure}   
   \figurenum{6}
   \centerline{\psfig{figure=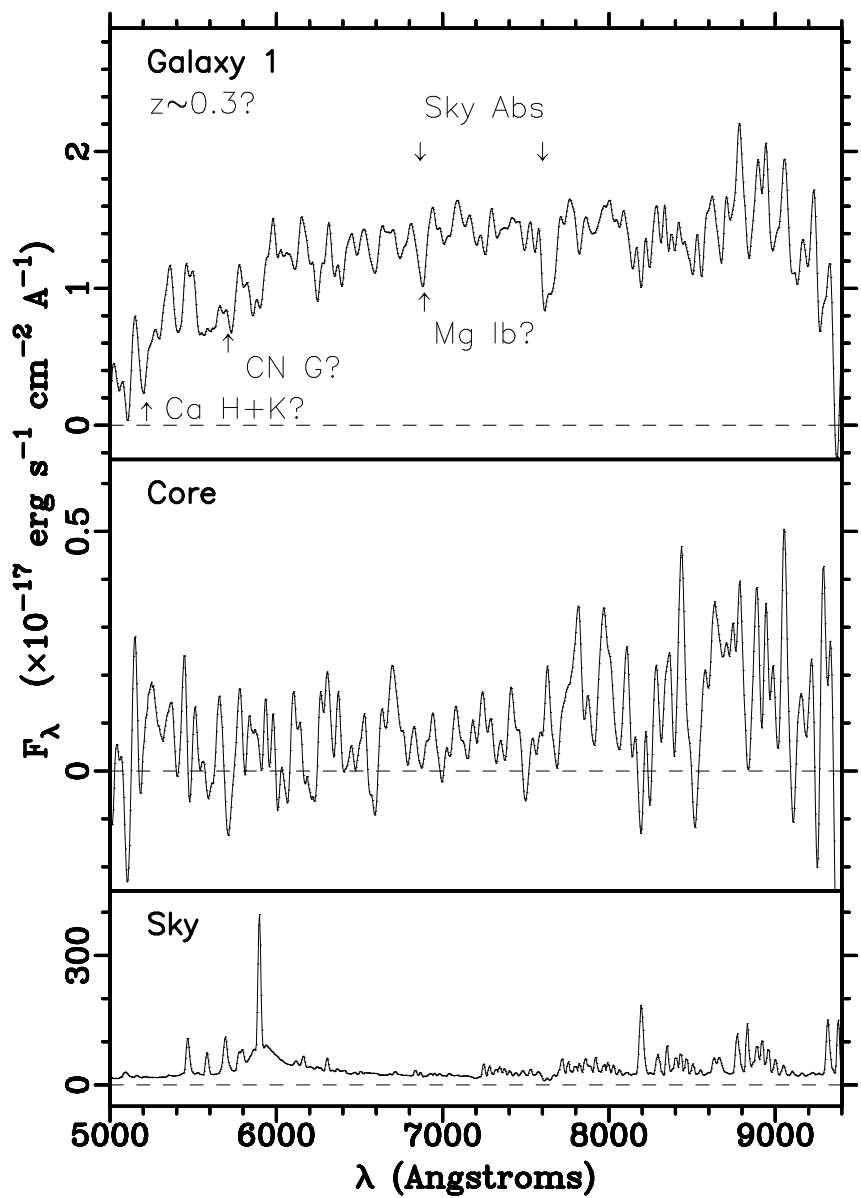,height=7in}}
   \caption{
      Optical spectra of the J0823+3096 system from a 1.2\,hr observation
      (s00) on the Palomar 5\,m telescope.
      Spectra are presented for the Galaxy counterpart to the radio lobe,
      the radio core counterpart, and the sky level during the observation.
      The target spectra have been smoothed to $\sim$20\,\AA\ resolution (FWHM). 
      Possible spectral features at $z$=0.33 are shown for Galaxy~1, 
      based mostly on the continuum shape. 
      The core was barely detected. 
      }
   \end{figure}

\begin{figure}   
   \figurenum{7}
   \centerline{\psfig{figure=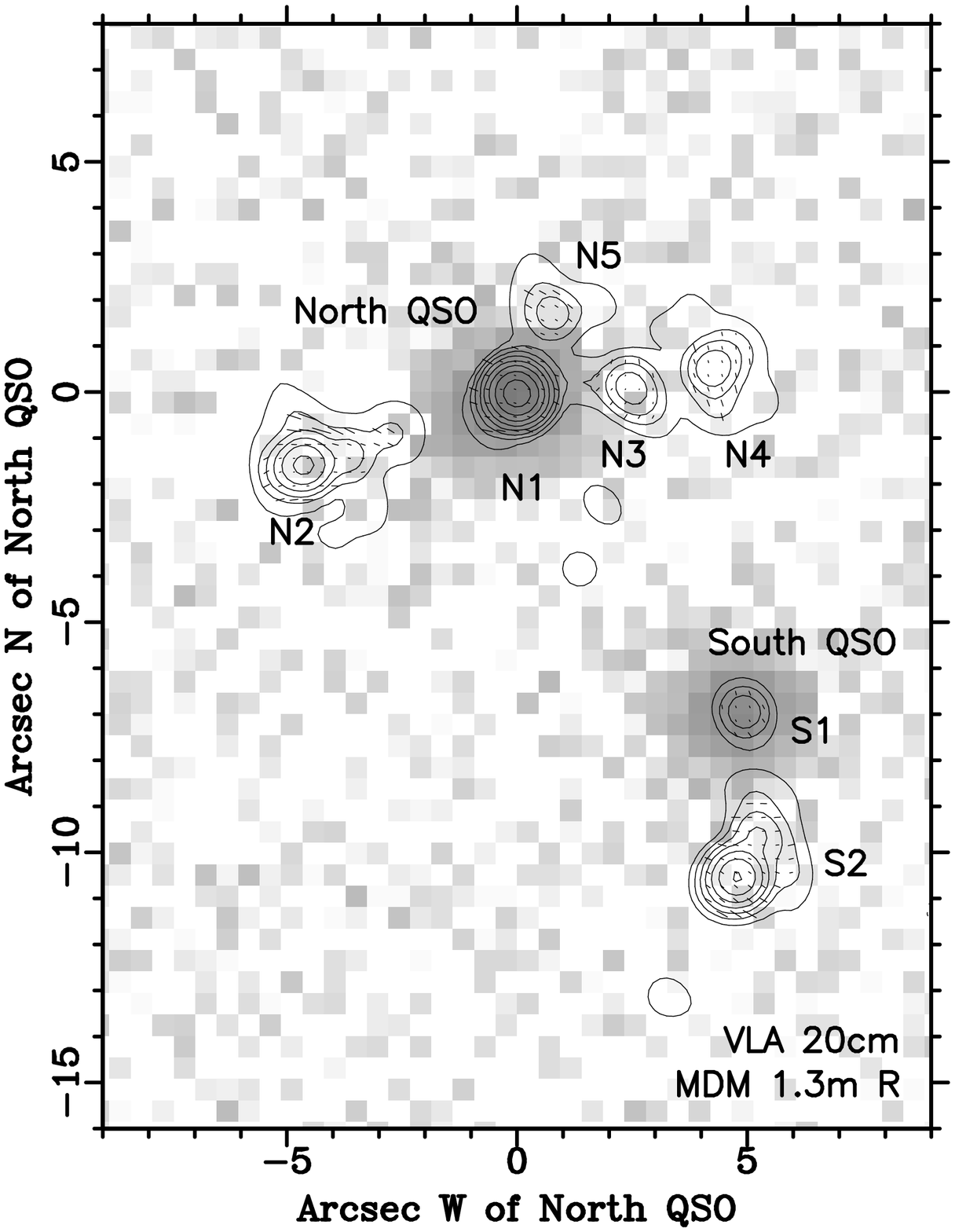,height=7in}}
   \caption{VLA A-array map of J0958+2947 at 20\,cm (r90b, 15\,min)
      with radio components labeled N1--N5 and S1--S2.
      The optical field (o90) is in logarithmic grey~scale, 
      aligned using the Northern QSO.
      The radio map is super-resolved to match the 0$\farcs$7 resolution 
      of the 6\,cm observation (Figure~2),
      and contours double from twice the off-source rms noise 
      of 0.21\,mJy\,beam$^{-1}$.  
      Fractional polarization vectors are parallel to the electric field,
      and are scaled so that the vector spacing corresponds to 33\% per beam.
      }
   \end{figure}

\begin{figure}   
   \figurenum{8}
   \centerline{\psfig{figure=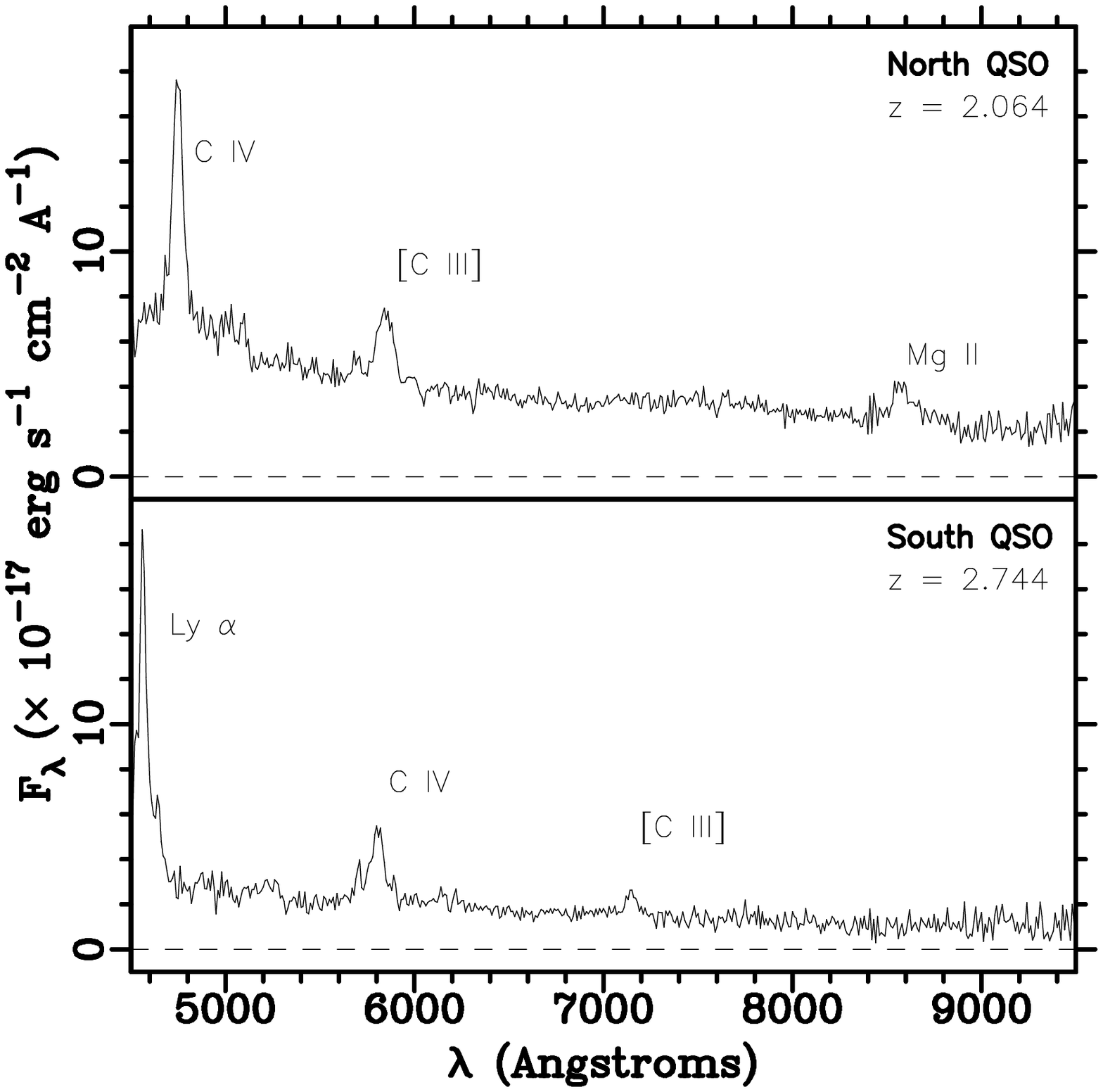,height=6in}}
   \caption{
      Optical spectra for J0958+2947 from a 10\,min 
      Palomar 5\,m observation (s90).
      Both objects show strong emission lines corresponding 
      to high redshift quasars.
      }
   \end{figure}

\begin{figure}   
   \figurenum{9}
   \centerline{\psfig{figure=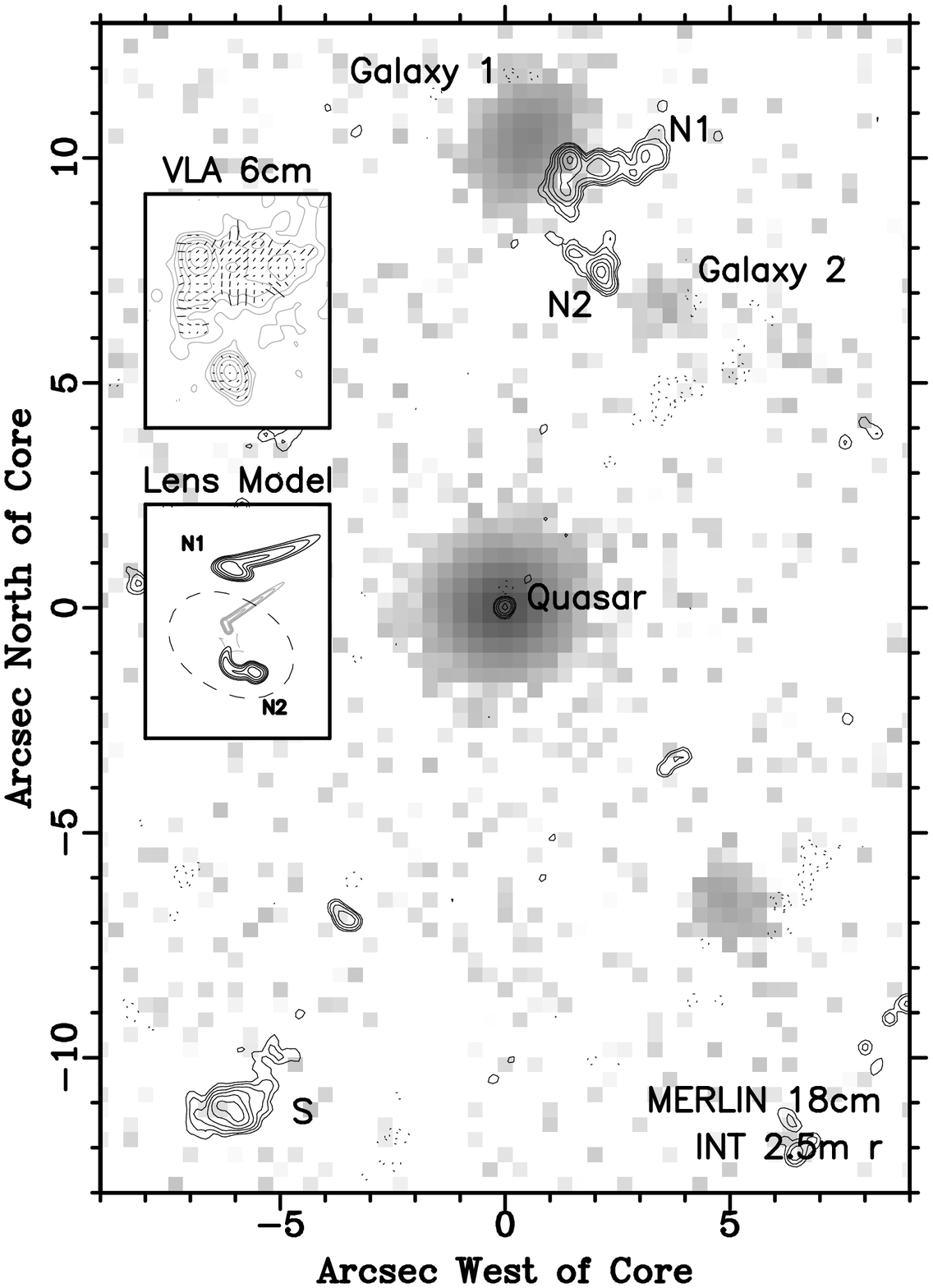,height=7in}}
   \caption{
      MERLIN map of J1622+3531 (r98a, 4\,hr) at 18\,cm. 
      The optical field (o99b) is shown in logarithmic grey~scale,
      aligned using the Quasar and radio core.
      Radio contours double from twice the off-source rms noise 
      of 54\,$\mu$Jy\,beam$^{-1}$, for a beam size of 0\farcs25 (FWHM). 
      The upper inset shows the VLA 6\,cm map (r98b) of the northern radio lobe
      at $0\farcs35$ resolution (rms noise 82\,$\mu$Jy\,beam$^{-1}$),
      with fractional polarization vectors scaled so that 33\% per beam
      corresponds to the vector spacing. 
      The lower inset shows a singular isothermal ellipsoid lens model,
      with N1 and N2 as lensed images of each other. 
      The unlensed source contours and the caustic (dashed) are shown in grey, 
      and the lensed image contours and critical line (dashed) are in black.
      }
\end{figure}

\begin{figure}   
   \figurenum{10}
   \centerline{\psfig{figure=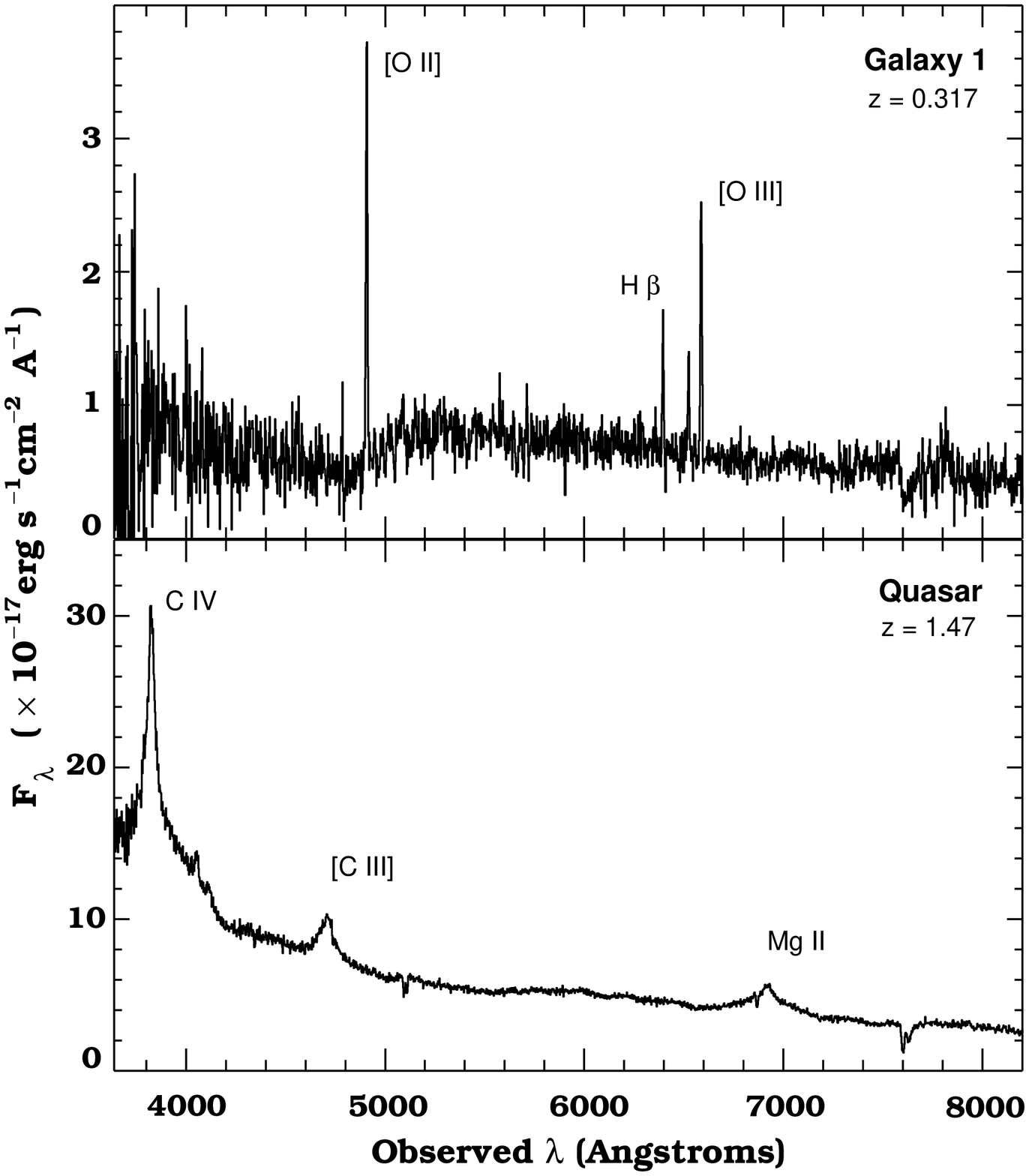,height=6in}}
   \caption{
      Optical spectra for J1622+3531, from a 15\,min observation
      using the Keck~II telescope (s99). 
      The radio core is a quasar at high redshift, 
      and the brighter lobe counterpart (Galaxy~1) 
      is a lower redshift star-forming galaxy. 
      }
\end{figure}

\begin{figure}
   \figurenum{11}
   \centerline{\psfig{figure=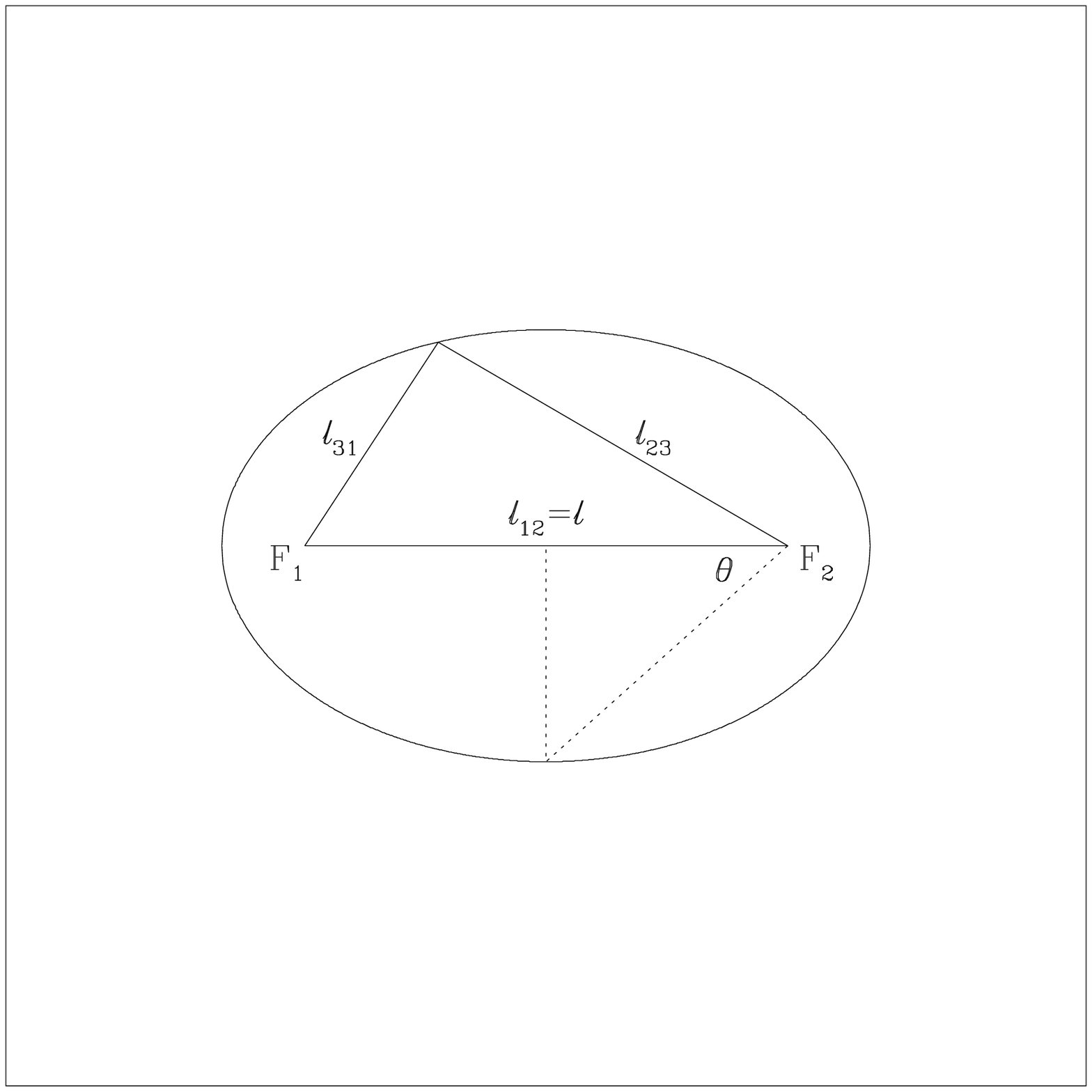,height=3.5in}}
   \caption{Definition of the configuration angle, $\theta$. 
      The triangle formed by three given radio components has sides
      $\ell=\ell_{12} > \ell_{23} > \ell_{31}$ and defines an ellipse with foci
      $F_1$ and $F_2$. The angle $\theta$ is given by
      $\theta = \arccos{\frac{\ell_{12}}{\ell_{23}+\ell_{31}}}$.
      }
\end{figure}


\begin{deluxetable}{lrrrl}
\tablenum{1}
\tablecolumns{5}
\tablecaption{Lens Candidate Optical Classification}
\tablehead{APM   & Lensed Lobe & FIRST & APM & \\
           Class & Candidates & Counterparts & Objects & Comments }
\startdata
   A+/A-- & 47\% & 55\% & 21\% & red, extended \\
   B+/B-- & 33\% & 40\% & 58\% & red, compact \\
   C+/C-- & 6\%  &  2\% & 10\% & blue, extended \\
   D+/D-- & 14\% &  3\% & 11\% & blue, compact \\
\enddata
\tablecomments{
   Here, `red' means that the APM $B$--$R${}$>$0.5\,mag,
   and `extended' means that the APM object classification was
   either `galaxy' or `merged', versus `stellar' or `noise-like'.
   The optical morphology was taken from the band
   in which the APM counterpart appeared brighter.
   The sign \mbox{(+/--)} indicates whether a separate counterpart to the 
   the presumed radio core was (present/absent).  APM statistics are 
   based on $\sim$13,400 $R${}$>$15 objects from 8 arbitrarily chosen 10$'$ fields.
   }
\end{deluxetable}

\begin{deluxetable}{ll}
\tablenum{2}
\tablecolumns{2}
\tablecaption{Lensed Lobe Candidate Selection Criteria}
\tablehead{Criterion & Comments}
\startdata
\sidehead{Radio Sources}
Component fluxes $>3$\,mJy
   & To ensure detection in radio follow-up. \\
Brightest $>10$\,mJy
   & To enable self-calibration in radio follow-up. \\
Separation $<60''$
   & Search radius to associate any 2 components. \\
Max.\ separation of $90''$
   & For triple and multiple FIRST sources. \\
Random assoc.\ prob.\ $<10\%$
   & To ensure significant component grouping. \\
Symmetry factor $S>0.5$
   & For triples, to eliminate chance associations. \\
Bending angle $\theta < 30^\circ$
   & For triples and multiples, to eliminate chance assoc. \\
Small Lobes
   & Reject if radio lobe size $>10''$, to ensure contrast. \\
\sidehead{Optical Counterparts}
Detected in \mbox{POSS-I}
   & Optical objects brighter than $R${}$\sim$20 or $B${}$\sim$22\,. \\
Coincides with lobe
   & Within 2\arcsec\ of the peak, or inside the FIRST FWHM ellipse. \\
Not likely host
   & Eliminate cases of one counterpart to both lobes in a double. \\
Not likely host
   & Reject if only one APM counterpart, on a compact component. \\
\sidehead{VLA sample}
APM close to lobe's peak
   & Reject if APM farther than $2''$ from fitted peak. \\
FIRST flux $>100$\,mJy
   & To distinguish lobes/cores in $<30$\,min (78 candidates). \\
Best visual ranking
   & To accommodate limited observing time (33 targets). \\
\enddata
\tablecomments{
  Before collapsing the FIRST sources, we eliminated VLA side-lobes, 
  and classified each FIRST component as compact 
  (major axis FWHM$<$2\arcsec) or extended. 
  We identified the ``core'' position of a collapsed source to be
  the centroid for doubles,
  or the middle component position for triples. 
  For multiples, we located the component closest to, 
  and within $\ell/4$ of, the centroid; or simply used the
  centroid if no components were close enough. 
  }
\end{deluxetable}

\begin{deluxetable}{ccrrl}
\def\tm#1{\tablenotemark{#1}}
\def\hh{\hphantom{a}}
\def\nd{\nodata~~~~~~~~~~~~~}
\tablenum{3}
\tablewidth{7in}
\tablecolumns{5}
\tablecaption{VLA Sample of Lensed Lobe Candidates}
\scriptsize
\tablehead{       FIRST            &   APM     &         VLA~~~~~~~~~&       CCD~~~~~~~~~~~~~~&\\
      J2000~Coordinates ~~mJy ~N&   R~~~~B~~~class& exp ~$\lambda$~ noise ~obs~~~& exp filt lim seeing obs & Notes}
\startdata
   07:12:29.7 +33:33:47 ~~214 ~2&  15.8  ~18.7 ~A$-$& ~9 ~6.0 ~0.33 ~{}r98b& \nd                              & not lobe\\     
   07:41:43.3 +33:35:39 ~~137 ~2&  18.4 ~$>$22 ~B$-$& 14 ~6.0 ~0.09 ~{}r98b& \nd                              & not lobe\\     
   08:09:13.7 +31:22:21 ~~176 ~2&  17.6  ~19.3 ~A$-$& ~8 ~6.0 ~0.13 ~{}r98b& \nd                              & not lobe\\     
   08:16:37.4 +50:03:39 ~~121 ~3&  18.7 ~$>$22 ~A$-$& 18 ~6.0 ~0.12 ~{}r98b& 55 ~$R$ ~25.9 ~$1\farcs8$ ~{}o00b& {\bf LENS}\\
   08:23:24.2 +39:06:26 ~~110 ~2&  19.4 ~$>$22 ~A$-$& 18 ~6.0 ~0.09 ~{}r98b& 80 ~$I$ ~27.2 ~$1\farcs6$ ~{}o00a& lens?\\
   08:24:59.4 +47:16:42 ~~137 ~2&  19.0  ~21.7 ~A$-$& 14 ~6.0 ~0.12 ~{}r98b& \nd                              & not lobe\\     
   08:28:22.7 +35:52:59 ~~146 ~3&  19.1  ~20.2 ~B$-$& 13 ~6.0 ~0.07 ~{}r98b& \nd                              & not lobe\\     
   08:49:26.9 +24:48:55 ~~134 ~2&  16.4  ~18.9 ~B$-$& ~9 ~6.0 ~0.07 ~{}r98b& 35 ~$R$ ~25.4 ~$1\farcs8$ ~{}o00b& not galaxy\\
   09:04:16.1 +42:38:04 ~1112 ~2&  19.9  ~21.6 ~B$-$& ~4 ~6.0 ~0.47 ~{}r98b& \nd                              & not lobe\\     
   09:04:54.8 +48:25:53 ~~114 ~2&  19.5  ~20.1 ~B$-$& 14 ~6.0 ~0.07 ~{}r98b& \nd                              & not lobe\\     
   09:58:58.8 +29:47:59 ~~138 ~2&  19.5  ~19.2 ~D$-$& 21 ~6.0 ~0.07 ~{}r90b& 20 ~$R$ ~24.2 ~$1\farcs7$ ~o90\hh& two QSOs\\
   10:36:42.0 +25:02:32 ~~119 ~2&  17.5  ~18.0 ~B$-$& 14 ~6.0 ~0.06 ~{}r98b& \nd                              & not lobe\\     
   11:10:43.2 +45:58:02 ~~178 ~2& $>$20  ~20.7 ~D$-$& ~9 ~6.0 ~0.09 ~{}r98b& \nd                              & not lobe\\     
   12:04:58.9 +49:40:26 ~~155 ~2&  17.6  ~21.1 ~A$-$& ~4 ~6.0 ~0.16 ~{}r98b& \nd                              & not lobe\\     
   12:21:19.0 +24:33:25 ~~362 ~2&  19.9 ~$>$22 ~B$-$& ~6 ~6.0 ~0.14 ~{}r98b& \nd                              & not lobe\\     
   12:21:27.6 +50:48:53 ~~218 ~2&  19.6  ~20.4 ~B$-$& ~9 ~6.0 ~0.08 ~{}r98b& \nd                              & not lobe\\     
   12:36:38.2 +28:40:30 ~~112 ~2&  18.3 ~$>$22 ~A$+$& 13 ~6.0 ~0.07 ~{}r98b& \nd                              & not lobe\\     
   13:20:05.6 +46:39:43 ~~102 ~3&  19.8 ~$>$22 ~B$-$& 14 ~6.0 ~0.10 ~{}r98b& 80 ~$R$ ~26.5 ~$1\farcs4$ ~{}o00b& not galaxy\\
   13:23:02.6 +29:41:33 ~~714 ~3& $>$20  ~21.2 ~D$-$& ~2 ~3.6 ~0.56 ~{}r90a& \nd                              & not lobe\\     
   13:37:07.6 +48:58:01 ~~163 ~2&  16.6  ~18.7 ~A$-$& ~9 ~6.0 ~0.18 ~{}r98b& \nd                              & not lobe\\     
   13:42:31.6 +35:07:11 ~~380 ~3&  19.9 ~$>$22 ~B$-$& ~5 ~6.0 ~0.24 ~r84\hh& 60 ~$I$ ~24.8 ~$1\farcs2$ ~{}o99a& not galaxy\\
   13:53:26.4 +57:25:50 ~~214 ~2&  19.0  ~20.9 ~B$-$& ~2 ~20. ~0.23 ~{}r91a& \nd                              & not lobe\\     
   13:57:30.5 +48:07:41 ~~158 ~2&  19.7 ~$>$22 ~A$+$& ~2 ~20. ~0.22 ~{}r91a& 60 ~$R$ ~26.4 ~$1\farcs3$ ~{}o00b& not galaxy\\
   14:17:58.4 +40:51:56 ~~105 ~2&  19.5  ~21.5 ~B$-$& 17 ~6.0 ~0.07 ~{}r98b& \nd                              & not lobe\\     
   14:19:46.0 +46:34:27 ~~141 ~2&  19.1  ~19.6 ~D$-$& ~2 ~20. ~0.24 ~{}r91a& \nd                                & not lobe\\     
   14:25:01.2 +30:27:14 ~~223 ~2& $>$20  ~22.0 ~D$+$& ~7 ~6.0 ~0.13 ~r95\hh& 80 ~$R$ ~25.8 ~$1\farcs3$ ~{}o00b& not galaxy\\
   15:28:41.9 +45:22:19 ~~153 ~2&  19.0  ~19.8 ~A$-$& 12 ~6.0 ~0.08 ~{}r98b& \nd                              & not lobe\\     
   15:49:12.7 +30:47:14 ~~877 ~2&  15.8  ~18.7 ~A$-$& ~1 ~6.0 ~0.48 ~{}r91b& ~5 ~$i$ ~25.6 ~$0\farcs9$ ~o91\hh& {\bf LENS}\\
   16:05:38.4 +54:39:23 ~~127 ~2& $>$20  ~21.4 ~D$+$& 25 ~6.0 ~0.08 ~{}r98b& 60 ~$I$ ~25.4 ~$1\farcs9$ ~{}o00b& not galaxy\\
   16:22:29.8 +35:31:27 ~~218 ~3&  19.9  ~21.7 ~B$+$& 18 ~6.0 ~0.15 ~{}r98b& ~5 ~$r$ ~27.0 ~$1\farcs0$ ~{}o99b& lens?\\
   16:41:57.4 +51:15:36 ~~138 ~2&  18.5  ~19.1 ~B$+$& 19 ~6.0 ~0.14 ~{}r98b& 55 ~$R$ ~25.7 ~$2\farcs2$ ~{}o00b& not lobe\\
   17:20:09.6 +38:25:41 ~~152 ~3&  18.9  ~21.4 ~B$-$&  6 ~6.0 ~0.13 ~{}r98b& \nd                              & not lobe\\     
   17:26:35.6 +32:13:22 ~~121 ~2&  17.0  ~17.6 ~B$-$& 17 ~6.0 ~0.10 ~{}r98b& \nd                              & not lobe\\
\enddata
\tablecomments{ \footnotesize
   The total radio flux density is given with the number of FIRST components.
   The reference coordinates refer to the likely radio core, when N=3, 
   otherwise to the radio components midpoint.
   The APM class is as described in Table~1. 
   Obs labels refer to Table~4, and exposure times are in minutes. 
   Noise estimates are off-source rms levels in mJy/beam for VLA,
   and 1$\sigma$ stellar detection limits for CCD observations.
   The notes summarize the outcome of this search, 
   where ``not galaxy'' means a stellar counterpart with no evidence
   for lensing, or a spurious APM detection. 
   }
\end{deluxetable}

\begin{deluxetable}{llll}
\def\tm#1{$^{\rm #1}$}
\tablenum{4}
\tablecolumns{4}
\tablecaption{Observations of the VLA Sample}
\scriptsize
\tablehead{ Label & Date & Description & Reference (Observer) }
\startdata
\sidehead{Radio Observations}
  r84&   1984.11.26&  VLA:A, 6\,cm, archival& (J.\ Condon) \\
  r96&   1996.12.18&  VLA:A, 3.6\&6\&20\,cm, archival& (S.\ Rawlings, AR365) \\
  r90a&  1990.05.03&  VLA:A, 3.6\,cm, archival& Burke\etal\ 1993 (AB568)\\
  r90b&  1990.05.07&  VLA:A, 6\&20\,cm, archival& Leh\'ar 1991 (AB456) \\
  r91a&  1991.06.28&  VLA:A, 20\,cm, archival& (S.\ Rawlings, AR250) \\
  r91b&  1991.08.02&  VLA:A, 6\,cm, archival& Leh\'ar\etal\ 1993 (AB611) \\
  r95&   1995.06.26&  VLA:A, 6\,cm, archival& (E.\ Falco, AF293) \\
  r98a&  1998.04.19$\rightarrow$20& \tm{a}MERLIN+Lovell, 18\,cm, 24\,hr& (J.\ Leh\'ar, MN/L/98A/33) \\
  r98b&  1998.05.26&  VLA:A, 6\,cm, 8\,hr& (J.\ Leh\'ar, AL434) \\
  r99&   1999.09.06&  VLA:A, 6\&20\,cm, 1\,hr& (J.\ Leh\'ar, AL434) \\
\sidehead{Optical CCD observations}
  o90&   1990.03.20&  \tm{b}MDM 1.3\,m, archival& Leh\'ar 1991 \\
  o91&   1991.09.04&  \tm{c}Palomar 5\,m, archival& Leh\'ar\etal\ 1993 \\
  o99a&  1999.01$\rightarrow$02& \tm{b}MDM 2.4\,m& (I.\ Yadigaroglu) \\
  o99b&  1999.07.18&  \tm{d}INT 2.5\,m& (M.\ Irwin) \\
  o00a&  2000.01.13&  \tm{e}FLW 1.2\,m& (E.\ Falco) \\
  o00b&  2000.02$\rightarrow$03& \tm{c}Palomar 1.5\,m& (A.\ Buchalter) \\
\sidehead{Optical Spectroscopy}
  s90&   1990.05.16&  \tm{c}Palomar 5\,m, 4-shooter, archival& (D.\ Schneider) \\
  s98&   1998.03.02&  \tm{d}INT 2.4\,m, IDS& (M.\ Irwin) \\
  s99&   1999.07.27&  \tm{f}Keck II 10\,m, LRIS& (R.\ Becker) \\
  s00&   2000.02.03&  \tm{c}Palomar 5\,m, Norris& (A.\ Buchalter) \\
\enddata
\tablenotetext{a}{\footnotesize
   MERLIN is operated by the University of Manchester as a
   national facility of the Particle Physics \& Astronomy Research
   Council (PPARC), in the United Kingdom.}
\tablenotetext{b}{\footnotesize
   The MDM Observatory is operated by a consortium of 
   the University of Michigan, Dartmouth College, and Columbia University. }
\tablenotetext{c}{\footnotesize
   The Palomar observatory is operated by the California Institute of Technology.}
\tablenotetext{d}{\footnotesize
   The Isaac Newton Telescope (INT) is operated on the island of La Palma by
   the Isaac Newton Group in the Spanish Observatorio del Roque de los Muchachos
   of the Instituto de Astrofisica de Canarias. }
\tablenotetext{e}{\footnotesize
   The Fred Lawrence Whipple Observatory on Mt Hopkins, AZ,
   is operated by the Smithsonian Astrophysical Observatory.  }
\tablenotetext{f}{\footnotesize
   The W.M.\ Keck Observatory is operated as a scientific
   partnership among the California Institute of Technology,
   the University of California and the National Aeronautics
   and Space Administration.  The Observatory was made possible
   by the generous financial support of the W.M.\ Keck Foundation.}.
\end{deluxetable}

\end{document}